\documentclass[preprint,review,3p,11pt]{elsarticle}

\usepackage[english]{babel}
\usepackage{amsmath}
\usepackage{amssymb}
\usepackage{amsthm}
\usepackage{amsfonts}
\usepackage{mathtools}
\usepackage[binary-units=true]{siunitx}
\usepackage{optidef}
\usepackage{blkarray}
\usepackage{algorithm}

\usepackage{subcaption}
\usepackage{graphicx}
\usepackage{xcolor}

\usepackage{booktabs}
\usepackage{tabularx}

\usepackage{tikz}
\usetikzlibrary{external,automata,trees,shadows,matrix,fit,decorations,positioning,shapes,plotmarks,patterns,arrows,calc,quotes,tikzmark,arrows.meta}

\newcommand{\mat}[1]{\ensuremath{{\mathbf{\MakeUppercase{#1}}}}}

\makeatletter
\renewcommand{\vec}[1]{%
	\ifcat\relax\noexpand#1%
	\ensuremath{\boldsymbol{\lowercase{#1}}}%
	\else
	\ensuremath{\mathbf{\lowercase{#1}}}%
	\fi
}
\makeatother
\newcommand{\trace}[1]{\ensuremath{\text{Tr}\left(#1\right)}}

\newcommand{\kron}{\ensuremath{\otimes}}

\newcommand{\ev}[1]{\ensuremath{\mathbb{E}\{{#1}\}}}
\newcommand{\sumlim}[2]{\ensuremath{\sum\limits_{#1}^{#2}}}
\makeatother

\makeatletter



\renewcommand{\max}[1]{\ensuremath{\underset{#1}{\text{max }}}}

\newcommand{\transpose}[1]{\ensuremath{{#1}^{\textsc{t}}}}

\newcommand{\R}{\ensuremath{\mathbb{R}}}
\newcommand{\norm}[1]{\left|\left|#1\right|\right|}

\usepackage{pifont}

\newcommand{\bigO}{\mathcal{O}}
\usepackage{multirow}
\usepackage{algcompatible}
\usepackage{adjustbox}

\makeatletter
\def\ps@pprintTitle{%
  \let\@oddhead\@empty
  \let\@evenhead\@empty
  \let\@oddfoot\@empty
  \let\@evenfoot\@oddfoot
}
\makeatother

\floatname{algorithm}{Algorithm}
\newcommand{\KwIn}[1]{\textbf{Input:} #1}
\newcommand{\KwOut}[1]{\textbf{Output:} #1}

\definecolor{r}{rgb}{0.63500,0.07800,0.18400}%
\definecolor{g}{rgb}{0.4660,0.6740,0.1880}%
\usepackage{hyperref}

\begin{document}

\begin{frontmatter}
		
% *** PAPER METADATA ***
\title{Grouped Variable Selection for Generalized Eigenvalue Problems}
		
\author[1,2]{Jonathan Dan\corref{cor1}\fnref{fn1,fn2}}\ead{jonathan.dan@esat.kuleuven.be}
\author[1,3]{Simon Geirnaert\fnref{fn1,fn2}}\ead{simon.geirnaert@esat.kuleuven.be}
\author[1]{Alexander Bertrand\fnref{fn2}}\ead{alexander.bertrand@esat.kuleuven.be}
		
\address[1]{KU Leuven, Department of Electrical Engineering (ESAT), STADIUS Center for Dynamical Systems, Signal Processing and Data Analytics, Kasteelpark Arenberg 10, 3001 Leuven, Belgium}
\address[2]{Byteflies, Borsbeeksebrug 22, 2600 Berchem, Belgium}
\address[3]{KU Leuven, Department of Neurosciences, Research Group ExpORL, Herestraat 49 box 721, 3000 Leuven, Belgium \\ \vspace*{-0.75cm}
}
		
\cortext[cor1]{Corresponding author}
\fntext[fn1]{These authors contributed equally.}
\fntext[fn2]{J. D., S. G., and A. B. are also with Leuven.AI - KU Leuven Institute for AI, 3000, Leuven, Belgium.}

% *** ABSTRACT ***
\begin{abstract}
	Many problems require the selection of a subset of variables from a full set of optimization variables. The computational complexity of an exhaustive search over all possible subsets of variables is, however, prohibitively expensive, necessitating more efficient but potentially suboptimal search strategies. We focus on sparse variable selection for generalized Rayleigh quotient optimization and generalized eigenvalue problems. Such problems often arise in the signal processing field, e.g., in the design of optimal data-driven filters. We extend and generalize existing work on convex optimization-based variable selection using semidefinite relaxations toward group-sparse variable selection using the $\ell_{1,\infty}$-norm. This group-sparsity allows, for instance, to perform sensor selection for spatio-temporal (instead of purely spatial) filters, and to select variables based on multiple generalized eigenvectors instead of only the dominant one. Furthermore, we extensively compare our method to state-of-the-art methods for sensor selection for spatio-temporal filter design in a simulated sensor network setting. The results show both the proposed algorithm and backward greedy selection method best approximate the exhaustive solution. However, the backward greedy selection has more specific failure cases, in particular for ill-conditioned covariance matrices. As such, the proposed algorithm is the most robust currently available method for group-sparse variable selection in generalized eigenvalue problems.
\end{abstract}

% *** RESEARCH HIGHLIGHTS AND KEYWORDS ***
		
	
\end{frontmatter}

% *** INTRODUCTION ***
\section{Introduction}
\label{sec:introduction}
\noindent
Variable selection is an important problem occurring in many mathematical engineering fields. Its goal is to select the subset of variables - often corresponding to specific sensor signals or features thereof - that have the largest impact on the optimization of a specific objective function. Such methods are often used, e.g., to identify the most relevant sensor nodes in a sensor network, or to find the optimal positions to place sensors in a predefined grid~\cite{chepuri2015sparsity}. These sensor selection problems arise in many signal processing-related fields, including telecommunication, where antenna placement is critical to the good functioning of a communication network~\cite{hamza2020sparse, gao2018sparse, wanlu2019controllable, hamza2018optimum, hamza2019sparse, hamza2020sparseConf, hamza2021sparse, zhai2021cognitive}, biomedical sensor arrays, e.g., in the context of electro-encephalography (EEG) channel selection or optimal positioning of wearable sensors~\cite{alotaiby2015review,narayanan2020analysis,narayanan2020optimal,dan2020efficient}, or wireless acoustic sensor networks, where a microphone subset needs to be selected~\cite{bertrand2011applications,zhang2018microphone}. The number of sensors is typically constrained by practical factors such as fabrication cost, bandwidth, or physical setup limitations, necessitating an appropriate selection of a limited number of sensors and their location.

In many signal processing applications, the objective function can be written as a generalized Rayleigh quotient (GRQ), which corresponds to solving a generalized eigenvalue decomposition (GEVD). Such GRQ- or GEVD-based objectives are encountered in various beamformer or filter design problems, for example, to maximize the signal-to-noise ratio (SNR)~\cite{mehanna2013joint,hamza2019hybrid,wanlu2019controllable,dan2020efficient}, or to maximize discriminative properties of the output signals of a filterbank, e.g., in biomedical sensor arrays~\cite{geirnaert2020fast,blankertz2007optimizing}. In these contexts, variable/sensor selection helps to reduce the computational complexity and power requirements of processing pipelines, to reduce the risk of overfitting of models, and to improve the overall setup.

In this paper, we focus on grouped variable selection in GRQ/GEVD problems, where the goal is to select a subset of predefined \textit{groups} of variables. For illustrative purposes, but without loss of generality, we will introduce the problem in the context of sensor selection for data-driven spatio-temporal filter design. In sensor networks, an intuitive grouping of the optimization variables is based on the finite impulse response (FIR) filter tap weights in each sensor. However, various (other) types of groupings exist, such as a grouped selection across different filterbands or output filters. Note that all presented methods are besides sensor networks applicable to any other application containing (group-sparse) variable selection for GRQ optimization and GEVD problems.

In sensor selection, the goal is to identify the optimal subset of $M$ out of $C$ available sensors where the choice of $M$ typically leads to a tradeoff between the optimization objective and satisfying some practical constraints. The exhaustive evaluation of all possible sensor combinations is a computationally costly operation. The selection of $M$ out of $C$ sensors is of combinatorial complexity $\frac{C!}{M!\,(C-M)!}$, where each evaluation requires a new GEVD computation, which in itself has a computational cost of $\bigO(M^3)$. Therefore, computationally efficient methods are required to solve the sensor selection problem. Two popular heuristic methods are found in the greedy forward selection (FS) and backward elimination (BE) algorithms~\cite{alotaiby2015review,dash1997feature}, which are easily applied to many selection problems, including the GEVD problem. However, their greedy nature strongly reduces the combinatorial exploration space, which can result in a highly suboptimal selection. Other approaches take the specific problem structure into account and combine optimization of the objective (e.g., the GRQ) with finding a sparse set of sensors. A specific subclass among these optimization-based approaches relaxes the sensor selection problem to a \emph{convex} optimization problem, which can be solved with off-the-shelf convex optimization solvers~\cite{joshi2009sensor}. More specifically for the GEVD problem, \cite{hamza2019hybrid} used the sparsity promoting $\ell_1$-norm for a purely spatial beamformer (see Section~\ref{sec:special-case-two-spatial}), which was extended in~\cite{hamza2020sparse} to a spatio-temporal beamformer using the $\ell_{1,\infty}$-norm as a group-sparse regularizer, albeit in a suboptimal manner (as we will show in Section~\ref{sec:special-case-one-filter}). Other approaches in radar beamforming employed the $\ell_{1,2}$-norm as a group-sparse regularizer in combination with successive convex approximation~\cite{hamza2020sparseConf,hamza2021sparse}. Furthermore, \cite{hamza2020sparse,hamza2020sparseConf,hamza2021sparse} only cover the case of a single output filter (multiple-input single-output (MISO) filtering), i.e., a single generalized eigenvector is computed, while several GEVD-based signal processing techniques, such as the common spatial patterns (CSP) filterbank, require the extraction of multiple eigenvectors (multiple-input multiple-output (MIMO) filtering).

The main contributions of this work are as follows:
\begin{itemize}
    \item We extend the GRQ/GEVD sensor selection for purely spatial filtering in~\cite{hamza2019hybrid} to spatio-\emph{temporal} filtering borrowing techniques from~\cite{mehanna2013joint}. This necessitates the use of a \emph{group-sparse} regularizer. When a sensor is eliminated, all corresponding filter lags should be put to zero.
    \item We add the possibility to take multiple filters (i.e., multiple generalized eigenvectors) into account (MIMO), whereas previous work only focused on the dominant generalized eigenvector (MISO). This requires consistent removal or zeroing of the filter coefficients corresponding to an eliminated sensor across \emph{all} filters. This approach can be employed in various other applications, where the notion of a shared selection exists.
    \item We provide an in-depth and statistical comparison of the proposed method with other state-of-the-art sensor selection methods in GEVD problems, which is largely missing in the aforementioned prior art.
\end{itemize}

The paper is structured as follows. First, the sensor selection problem for GEVD and GRQ optimization is introduced in Section~\ref{sec:sensor-selection-for-gevd-problems}. Next, the convex optimization-based group-sparse sensor selection is explained in Section \ref{sec:optimal-selection}. We then thoroughly compare the proposed method with other (benchmark) sensor selection methods on simulated data in Section~\ref{sec:benchmark-study}. In Section~\ref{sec:epilepsy}, we provide an example of the developed method applied on real-world data, in the context of mobile epileptic seizure monitoring. Finally, conclusions are drawn in Section~\ref{sec:conclusion}.

\subsection{Notation}
\label{sec:notations}
\noindent
Scalars, vectors, and matrices are denoted by a lowercase ($x$), bold lowercase ($\vec{x}$), and bold uppercase letter ($\mat{X}$). The element of matrix $\mat{X}$ on the $i$\textsuperscript{th} row and $j$\textsuperscript{th} column is given by $x_{ij}$. $\transpose{\mat{X}}$ denotes the transpose of a matrix $\mat{X}$ and $\trace{\mat{X}}$ denotes the trace of $\mat{X}$. The $N \times N$ identity matrix is denoted by $\mat{I}_N$, while $\mat{0}_N$ denotes an $N \times N$ matrix with zeros. The $\ell_{\infty}$-norm of a vector (i.e., the maximal absolute value) is written as $\norm{\vec{x}}_{\infty}$, the $\ell_1$-norm of a vector (i.e., the sum of the absolute values) as $\norm{\vec{x}}_1$, the $\ell_2$-norm of a vector (i.e., the square root of the sum of squared elements) as $\norm{\vec{x}}_2$, and the max-norm of a matrix (i.e., the maximal absolute value across all elements) as $\norm{\mat{X}}_{\textnormal{max}}$. $\mat{X} \succcurlyeq 0$ denotes that $\mat{X}$ is a positive semidefinite matrix. The Kronecker-delta is written as $\delta_{ij}$ (i.e., $\delta_{ij} = 0 \textnormal{ if } i \neq j ; ~ \delta_{ij} = 1 \textnormal{ if } i = j$). Finally, the Kronecker-product of matrices $\mat{X} \in \R^{I_x \times J_x}$ and $\mat{Y} \in \R^{I_y \times J_y}$ is defined as:
\[
\mat{X} \kron \mat{Y} = \begin{bmatrix} x_{11}\mat{Y} & \cdots & x_{1J_x}\mat{Y} \\
\vdots & \ddots & \vdots \\
x_{I_x1}\mat{Y} & \cdots & x_{I_xJ_x}\mat{Y}
\end{bmatrix} \in \R^{I_xI_y \times J_xJ_y}.
\]

% *** METHODS ***
\section{Sensor selection for GEVD problems}
\label{sec:sensor-selection-for-gevd-problems}
\noindent
Consider a setting with $C$ sensors and two stationary zero-mean multi-sensor signals $\vec{x}_1(t)\in \R^{CL}$ and $\vec{x}_2(t)\in \R^{CL}$, where $t$ denotes the sample (time) index and $L$ denotes the group size as explained below. $\vec{x}_1(t)$ and $\vec{x}_2(t)$ could represent the $C$ sensor signals measured during two different states (e.g., EEG during movement of the left arm and movement of the right arm~\cite{blankertz2007optimizing}), or they could represent two signal components that are both simultaneously present in the sensor signals (e.g., target signal and noise). We assume that the entries of $\vec{x}_1(t)$ and $\vec{x}_2(t)$ are grouped in blocks of $L$ entries, each group corresponding to a single sensor. For example, a group could consist of $L$ frequency subbands or other features extracted from a single-sensor signal. For illustrative purposes, but without loss of generality, we focus here on spatio-temporal filter design, in which case the $L$ entries of a group correspond to a delay line of length $L$. In this case, the vector $\vec{x}_1(t)\in \R^{CL}$ can be represented as $\vec{x}_1(t) = \transpose{\begin{bmatrix} \transpose{\underline{\vec{x}}_{1,1}(t)} & \transpose{\underline{\vec{x}}_{1,2}(t)} & \cdots & \transpose{\underline{\vec{x}}_{1,C}(t)} \end{bmatrix}}$ where  $\underline{\vec{x}}_{1,c}(t) = \transpose{\begin{bmatrix} x_{1,c}(t) & x_{1,c}(t-1) & \cdots & x_{1,c}(t-L+1) \end{bmatrix}}$ represents the causal FIR filter taps corresponding to the $c$\textsuperscript{th} sensor (similarly for $\vec{x}_2(t)$). 

The goal is to find a spatio-temporal filter represented by $\vec{w} \in \R^{CL}$ which optimally discriminates between the two signals $\vec{x}_1(t)$ and $\vec{x}_2(t)$. Optimal discrimination corresponds to maximizing the energy of the output signal $y_1(t) = \transpose{\vec{w}}\vec{x}_1(t)$, while minimizing the energy of the output signal $y_2(t) = \transpose{\vec{w}}\vec{x}_2(t)$. The optimal $\vec{w}$ is thus found by maximizing:
\begin{maxi}|s|
{\substack{\vec{w} \in \R^{CL}}}{\frac{\ev{(\transpose{\vec{w}}\vec{x}_1(t))^2}}{\ev{(\transpose{\vec{w}}\vec{x}_2(t))^2}} = \frac{\transpose{\vec{w}}\mat{R}_1\vec{w}}{\transpose{\vec{w}}\mat{R}_2\vec{w}},}{\label{eq:rayleigh-quotient}}{}
\end{maxi}
where $\mat{R}_1 = \ev{\vec{x}_1(t)\transpose{\vec{x}_1(t)}} \in \R^{CL \times CL}$ and $\mat{R}_2= \ev{\vec{x}_2(t)\transpose{\vec{x}_2(t)}} \in \R^{CL \times CL}$ are the corresponding covariance matrices. Assuming ergodicity and given $T$ samples of the signals $\vec{x}_1(t)$ and $\vec{x}_2(t)$, these covariance matrices can be estimated as $\mat{R}_1 = \ev{\vec{x}_1(t)\transpose{\vec{x}_1(t)}} \approx \frac{1}{T} \sumlim{t = 0}{T-1}\vec{x}_1(t)\transpose{\vec{x}_1(t)}$ and similarly for $\mat{R}_2$. The problem in~\eqref{eq:rayleigh-quotient} is known as a generalized Rayleigh quotient (GRQ) optimization. In the case where  $\vec{x}_1(t)$ and $\vec{x}_2(t)$ represent the target signal and noise components, respectively, \eqref{eq:rayleigh-quotient} implies a maximization of the signal-to-noise ratio, resulting in a so-called max-SNR filter~\cite{vanveen1988beamforming}. In max-SNR filtering, the covariance matrices $\mat{R}_1$ and $\mat{R}_2$ thus correspond to the spatio-temporal covariance matrices related to the target signal and the noise, respectively. In the CSP framework~\cite{blankertz2007optimizing}, these covariance matrices correspond to the two signal classes that have to be discriminated (e.g., left versus right hand movement).

Because of the scale-invariance of $\vec{w}$ in~\eqref{eq:rayleigh-quotient}, we can arbitrarily set the output power depicted in the denominator to  $\transpose{\vec{w}}\mat{R}_2\vec{w} = 1$. Using the method of Lagrange multipliers to solve~\eqref{eq:rayleigh-quotient} then leads to a GEVD~\cite{yan2006trace}:
\[
\mat{R}_1\vec{w} = \lambda\mat{R}_2\vec{w}.
\]
The optimal filter $\vec{w}$ corresponds to the generalized eigenvector (GEVc) corresponding to the largest generalized eigenvalue (GEVl).

In various applications, the GRQ optimization of~\eqref{eq:rayleigh-quotient} for MISO filtering is generalized to MIMO filtering, i.e., multiple output filters. In this MIMO case, the goal is to find a filterbank of $K$ spatio-temporal filters $\mat{W} \in \R^{CL \times K}$ for which the sum of the energies of the multiple output signals is maximally discriminative:
\begin{maxi}|s|
{\substack{\mat{W} \in \R^{CL \times K}}}{\frac{\trace{\transpose{\mat{w}}\mat{R}_1\mat{w}}}{\trace{\transpose{\mat{w}}\mat{R}_2\mat{w}}}}{\label{eq:rayleigh-quotient-multiple-outputs}}{}
\addConstraint{\transpose{\mat{w}}\mat{R}_2\mat{w} = \mat{I}_{K},}
\end{maxi}
with $\trace{\cdot}$ denoting the trace operator and $\mat{I}_K$ the $K \times K$ identity matrix. The constraint in~\eqref{eq:rayleigh-quotient-multiple-outputs} ensures that the $K$ output channels are orthogonal to each other with respect to the signal component $\vec{x}_2(t)$. This constraint is added to obtain $K$ different filters. By plugging this constraint in the cost function in~\eqref{eq:rayleigh-quotient-multiple-outputs}, we obtain:
\begin{maxi}|s|
{\substack{\mat{W} \in \R^{CL \times K}}}{\trace{\transpose{\mat{w}}\mat{R}_1\mat{w}}}{\label{eq:rayleigh-quotient-multiple-outputs-reformulation}}{}
\addConstraint{\transpose{\mat{w}}\mat{R}_2\mat{w} = \mat{I}_{K}.}
\end{maxi}
The solution of~\eqref{eq:rayleigh-quotient-multiple-outputs-reformulation} is now given by taking the $K$ GEVcs corresponding to the $K$ largest GEVls:
\begin{equation}
\label{eq:gevd-multiple-outputs}
	\mat{R}_1\mat{W} = \mat{R}_2\mat{W}\mat{\Lambda}.
\end{equation}

This generalization to MIMO filters ($K > 1$) is crucial in classification tasks and discriminative analysis as in the CSP framework or in Fisher's discriminant analysis, where the data are projected into a $K$-dimensional feature space instead of a one-dimensional space. The number of output filters $K$ then introduces a tradeoff between how much information from the original data is preserved and the GRQ~\eqref{eq:rayleigh-quotient-multiple-outputs}, which becomes smaller (worse) for larger $K$.

Our goal is to find the optimal subset of $M \leq C$ out of $C$ sensors, with $M \geq K$, for which the ratio of traces in~\eqref{eq:rayleigh-quotient-multiple-outputs} is maximal. Note that eliminating a sensor means that all time lags corresponding to that sensor need to be zero. Furthermore, the selected sensors must be consistent across all $K$ filters (i.e., columns of $\mat{W}$) to be able to physically select only a few sensors. That is why a \emph{group-sparse} sensor selection is required, i.e., the filter weights are grouped per sensor across time lags and filters, and whole groups are put to zero (i.e., eliminated) rather than the individual elements in a group. In the next section, we present a convex optimization-based approach for this sensor selection problem.

\section{Optimal group-sparse sensor selection}
\label{sec:optimal-selection}
\noindent
In this section, we generalize the optimal sensor selection and array design method of~\cite{hamza2019hybrid}, which focuses on GRQ optimization and GEVD for purely spatial filtering (i.e., $L = 1$) and for MISO filtering (i.e., $K = 1$), to spatio-temporal filtering and MIMO filtering. Our derivation is based on a similar $\ell_{1,\infty}$-norm regularization technique as proposed in~\cite{mehanna2013joint} for multicast beamforming and antenna selection. It is noted that during the consolidation of this work, a similar idea to introduce group-sparsity in GEVD problems was published in~\cite{hamza2020sparse} in the meantime, independently from our work. The work in~\cite{hamza2020sparse} establishes the $L>1$ case, yet without generalizing to the $K>1$ case as also targeted here. Furthermore, our proposed generalization differs from~\cite{hamza2020sparse} on another crucial aspect, which will be pointed out throughout the derivation (see Section~\ref{sec:special-case-one-filter}), and which makes that~\cite{hamza2020sparse} can not be treated as a special case of our proposed general framework. In Section~\ref{sec:benchmark-study}, we will also empirically compare with~\cite{hamza2020sparse} for the $K=1$ setting and demonstrate the superiority of our generalization.

Before pursuing group sparsity in~\eqref{eq:rayleigh-quotient-multiple-outputs-reformulation}, let us first vectorize $\mat{W} \in \R^{CL \times K}$ as $\vec{w} \in \R^{CLK}$, with $\vec{w}_k, k \in \{1,\dots,K\}$, the $k$\textsuperscript{th} spatio-temporal filter:
\begin{equation}
\label{eq:w-definition}
\vec{w} = \begin{bmatrix}
	\vec{w}_1 \\
	\vdots \\
	\vec{w}_K \\
\end{bmatrix}, \vec{w}_k = \begin{bmatrix}
	\vec{w}_{k,1} \\
	\vdots \\
	\vec{w}_{k,C} \\
\end{bmatrix}, \textnormal{ and } \vec{w}_{k,c} = \begin{bmatrix}
	w_{k,c,1} \\
	\vdots \\
	w_{k,c,L} \\
\end{bmatrix}.
\end{equation}
The optimization problem in~\eqref{eq:rayleigh-quotient-multiple-outputs-reformulation} then becomes:
\begin{mini}|s|
{\substack{\vec{w} \in \R^{CLK}}}{\transpose{\vec{w}}\left(\mat{I}_K\kron\mat{R}_2\right)\vec{w}}{\label{eq:vectorized-optimization-problem}}{}
\addConstraint{\transpose{\vec{w}}_k\mat{R}_1\vec{w}_{k'}}{= \delta_{kk'},\quad}{\forall k,k' \in \{1,\dots,K\},}
\end{mini}
with $\kron$ the Kronecker-product and $\delta_{kk'}$ the Kronecker-delta (i.e., $\delta_{kk'} = 0 , ~ \forall k \neq k' ; ~ \delta_{kk'} = 1 , ~ \forall k = k'$). Note that we changed the problem in~\eqref{eq:rayleigh-quotient-multiple-outputs-reformulation} to a minimization problem to accommodate for an easy introduction of the sparse regularization term. It can be shown that the solution of~\eqref{eq:rayleigh-quotient-multiple-outputs-reformulation} and~\eqref{eq:vectorized-optimization-problem} are the same up to an arbitrary scaling on each $\vec{w}_k$, which is irrelevant as generalized eigenvectors are defined up to a scaling. Using the filter-selector matrix $\mat{S}_k \in \R^{CL \times CLK}$, where the subscript indicates the selected coefficients of $\vec{w}$:
\begin{equation*}
  \begin{blockarray}{*{8}{c}}
    \begin{block}{*{8}{>{$\footnotesize}c<{$}}}
      & 1 &  & k-1 & k & k+1 &  & K\\
    \end{block}
    \begin{block}{*{1}{c}[*{7}{c}]}
      \mat{S}_k = & \mat{0}_{CL} & \dots & \mat{0}_{CL} & \mat{I}_{CL} & \mat{0}_{CL} & \dots & \mat{0}_{CL} \\
    \end{block}
  \end{blockarray}
\end{equation*}
with an identity matrix on the $k$\textsuperscript{th} position to select the $k$\textsuperscript{th} filter $\vec{w}_k$ from $\vec{w}$, i.e., $\mat{S}_k\vec{w}=\vec{w}_k$, \eqref{eq:vectorized-optimization-problem} can be rewritten as:
\begin{mini}|s|
{\substack{\vec{w} \in \R^{CLK}}}{\transpose{\vec{w}}\left(\mat{I}_K\kron\mat{R}_2\right)\vec{w}}{\label{eq:vectorized-optimization-problem-with-selector-matrix}}{}
\addConstraint{\transpose{\vec{w}}\transpose{\mat{S}}_k\mat{R}_1\mat{S}_{k'}\vec{w}}{= \delta_{kk'},\quad}{\forall k,k' \in \{1,\dots,K\}.}
\end{mini}

\subsection{Group-sparsity promoting regularization}
\label{sec:group-sparsity}
\noindent
The goal is now to introduce sparsity in~\eqref{eq:vectorized-optimization-problem} on the \emph{sensor} level. This sparsity on the sensor level corresponds to a group-sparse constraint on~\eqref{eq:vectorized-optimization-problem}, as all lags of all output filters corresponding to a particular sensor need to be set to zero. Therefore, as in~\cite{mehanna2013joint}, we deploy the convex sparsity-promoting $\ell_{1,\infty}$-norm as a proxy for the optimal but non-convex $\ell_0$-norm as a regularization term in~\eqref{eq:vectorized-optimization-problem}. Note that using $\ell_{1,\infty}$-norm as a sparsity-promoting norm on the filter weights themselves only requires the weights to be zero or non-zero. In the case where the entries of $\vec{w}$ would be constrained to be binary variables, one could also consider enforcing binary sparsity, which has already been employed, for example, in the context of direction-of-arrival estimation~\cite{wang2014reconfigurable,wang2015adaptive} (this, however, does not incorporate the optimization of the filter weights as desired in our problem statement~\eqref{eq:rayleigh-quotient-multiple-outputs}). 

To simplify the notations in the remainder of the derivations, we define the permutation matrix $\mat{P} \in \R^{CLK \times CLK}$ that permutes the elements of $\vec{w}$ such that they are first ordered by sensor and then by filter and lags (instead of first by filter as in~\eqref{eq:w-definition}), resulting in $\Tilde{\vec{w}}$:
\begin{equation}
\label{eq:wtilde-definition}
    \Tilde{\vec{w}} = \mat{P}\vec{w} = \begin{bmatrix}
	\Tilde{\vec{w}}_1 \\
	\vdots \\
	\Tilde{\vec{w}}_C \\
\end{bmatrix} \textnormal{ and } \Tilde{\vec{w}}_c = \begin{bmatrix}
	\vec{w}_{1,c} \\
	\vdots \\
	\vec{w}_{K,c} \\
\end{bmatrix}.
\end{equation}
Using this notation, the $\ell_{1,\infty}$-norm on the sensor level is defined as:
\begin{equation}
    \norm{\vec{w}}_{1,\infty} = \sumlim{c = 1}{C} \norm{\Tilde{\vec{w}}_{c}}_{\infty} = \sumlim{c = 1}{C} \max{k = 1,\dots,K} \;\norm{\vec{w}_{k,c}}_{\infty},
\end{equation}
where $\norm{\Tilde{\vec{w}}_{c}}_{\infty}$ corresponds to the maximal absolute value across all lags and filters corresponding to sensor $c$. As the $\ell_1$-norm induces sparsity, while the $\ell_{\infty}$-norm is only zero when all elements are zero, the $\ell_{1,\infty}$-norm can be used to put groups of coefficients across lags and filters corresponding to one sensor to zero. Furthermore, in~\cite{mehanna2013joint}, it is also shown that any sparsity-inducing norm can be replaced with the squared norm without changing the regularization properties of the problem. Therefore, the sensor selection problem with the group-sparse regularization term becomes:
\begin{mini}|s|
{\substack{\vec{w} \in \R^{CLK}}}{\transpose{\vec{w}}\left(\mat{I}_K\kron\mat{R}_2\right)\vec{w} + \mu\left(\sumlim{c = 1}{C} \norm{\Tilde{\vec{w}}_{c}}_{\infty}\right)^2}{\label{eq:group-sparse-optimization-problem}}{}
\addConstraint{\transpose{\vec{w}}\transpose{\mat{S}}_k\mat{R}_1\mat{S}_{k'}\vec{w}}{= \delta_{kk'},\quad}{\forall k,k' \in \{1,\dots,K\},}
\end{mini}
where the regularization parameter $\mu$ can be used to control the solution's sparsity and thus the number of sensors selected. Note that this is not yet a convex optimization problem due to the quadratic equality constraints.

\subsection{Semidefinite formulation and relaxation}
\label{sec:semidefinite-formulation}
\noindent
To transform~\eqref{eq:group-sparse-optimization-problem} into a convex semidefinite problem (SDP), we are using the following trick, as suggested in~\cite{luo2010semidefinite,mehanna2013joint,hamza2019hybrid}:
\begin{eqnarray*}
\transpose{\vec{w}}\left(\mat{I}_K\kron\mat{R}_2\right)\vec{w}
& = & \trace{\transpose{\vec{w}}\left(\mat{I}_K\kron\mat{R}_2\right)\vec{w}} \\  
& = & \trace{\left(\mat{I}_K\kron\mat{R}_2\right)\vec{w}\transpose{\vec{w}}} \\
& = & \trace{\left(\mat{I}_K\kron\mat{R}_2\right)\mat{V}},
\end{eqnarray*}
where the second equality holds because of the cyclic property of the trace. Per definition, $\mat{V} = \vec{w}\transpose{\vec{w}} \in \R^{CLK \times CLK}$ is thus a rank-1 positive semidefinite matrix. Similarly, the equality constraints can be reformulated as:
\begin{equation*}
    \trace{\mat{R}_1\mat{S}_{k'}\mat{V}\transpose{\mat{S}}_k} = \delta_{kk'}, \forall k,k' \in \{1,\dots,K\}.
\end{equation*}

Using the following definition of $\Tilde{\mat{V}}$:
\begin{equation*}
    \Tilde{\mat{V}} = \tilde{\vec{w}}\transpose{\tilde{\vec{w}}} = \mat{P}\mat{V}\transpose{\mat{P}} = \begin{bmatrix}
	\tilde{\mat{V}}_{11} &  \cdots & \tilde{\mat{V}}_{1C} \\
	\vdots & \ddots & \vdots\\
	\tilde{\mat{V}}_{C1} &  \cdots & \tilde{\mat{V}}_{CC}\\
\end{bmatrix},
\end{equation*}
the group-sparse regularization term in~\eqref{eq:group-sparse-optimization-problem} can be reformulated similarly to~\cite{mehanna2013joint}:
\begin{eqnarray}
\left(\sumlim{c = 1}{C} \norm{\Tilde{\vec{w}}_{c}}_{\infty}\right)^2
& = & \sumlim{c_1=1}{C}\sumlim{c_2=1}{C}\norm{\Tilde{\vec{w}}_{c_1}}_{\infty}\norm{\Tilde{\vec{w}}_{c_2}}_{\infty} \nonumber\\  
& = & \sumlim{c_1=1}{C}\sumlim{c_2=1}{C}\norm{\tilde{\mat{V}}_{c_1c_2}}_{\textnormal{max}} \nonumber\\
& = & \trace{\vec{1}_C\transpose{\vec{1}}_C\mat{U}}, \label{eq:U-introduction}
\end{eqnarray}
where the max-norm $\norm{\mat{A}}_{\textnormal{max}} = \max{i,j} |a_{ij}|$ is the elementwise maximum over a matrix, $\vec{1}_C \in \R^C$ denotes an all-ones vector of length $C$, and where $\mat{U} \in \R^{C \times C}$ is equal to:
\begin{equation}
\label{eq:U-definition}
    \mat{U} = \begin{bmatrix}
	\norm{\tilde{\mat{V}}_{11}}_{\textnormal{max}} &  \cdots & \norm{\tilde{\mat{V}}_{1C}}_{\textnormal{max}} \\
	\vdots & \ddots & \vdots\\
	\norm{\tilde{\mat{V}}_{C1}}_{\textnormal{max}} &  \cdots & \norm{\tilde{\mat{V}}_{CC}}_{\textnormal{max}}\\
\end{bmatrix}.
\end{equation}
Using the definition of $\mat{U}$ in~\eqref{eq:U-definition}, we finally obtain:
\begin{mini}|s|
{\substack{\mat{V} \in \R^{CLK\times CLK}, \\ \mat{U} \in \R^{C \times C}}}{\trace{\left(\mat{I}_K\kron\mat{R}_2\right)\mat{V}} + \mu\trace{\vec{1}_C\transpose{\vec{1}}_C\mat{U}}}{\label{eq:sdrWithRank}}{}
\addConstraint{\trace{\mat{R}_1\mat{S}_{k'}\mat{V}\transpose{\mat{S}}_k} = \delta_{kk'}, \forall k,k' \in \{1,\dots,K\}}
\addConstraint{\mat{U} \geq |\mat{S}_{k,l}\mat{V}\transpose{\mat{S}}_{k',l'}|, \forall k,k' \in \{1,\dots,K\}}
\addConstraint{}{}{\text{and }\forall l,l' \in \{1,\dots,L\}}
\addConstraint{\mat{V} \succcurlyeq 0, \textnormal{rank}(\mat{V}) = 1,}
\end{mini}
with the selector-matrix $\mat{S}_{k,l} \in \R^{C \times CLK}$ selecting all coefficients across $C$ sensors for a particular filter $k$ and lag $l$. The second constraint is an element-wise inequality, which ensures that each element of $\mat{U}$ (i.e., for each pair of sensors) is larger than the corresponding element for the corresponding pair of sensors across all filters and lags (expressed by the $\forall$ over the filter and lag indices), and thus implements the max-norm operation. The last two constraints ensure the equivalence between $\mat{V}$ and $\vec{w}\transpose{\vec{w}}$.

However,~\eqref{eq:sdrWithRank} is still not a convex optimization problem due to the rank-1 constraint. Therefore, we approximate~\eqref{eq:sdrWithRank} by relaxing the rank constraint, which is a technique known as semidefinite relaxation (SDR) and results in an SDP~\cite{luo2010semidefinite}:
\begin{mini}|s|
{\substack{\mat{V} \in \R^{CLK\times CLK}, \\ \mat{U} \in \R^{C \times C}}}{\trace{\left(\mat{I}_K\kron\mat{R}_2\right)\mat{V}} + \mu\trace{\vec{1}_C\transpose{\vec{1}}_C\mat{U}}}{\label{eq:sdr}}{}
\addConstraint{\trace{\mat{R}_1\mat{S}_{k'}\mat{V}\transpose{\mat{S}}_k} = \delta_{kk'}, \forall k,k' \in \{1,\dots,K\}}
\addConstraint{\mat{U} \geq |\mat{S}_{k,l}\mat{V}\transpose{\mat{S}}_{k',l'}|, \forall k,k' \in \{1,\dots,K\}}
\addConstraint{}{}{\quad\quad\quad\quad\quad\quad \text{and }\forall l,l' \in \{1,\dots,L\}}
\addConstraint{\mat{V} \succcurlyeq 0.}
\end{mini}
This SDR results in practice in a good approximation of the underlying rank-1 solution, potentially using a post-hoc rank-1 approximation of the solution. Note, however, that we are here not interested in the optimal filter coefficients themselves, but only in the selected sensors, which can be retrieved as the non-zero elements of the diagonal of $\mat{U}$. Typically, the GEVD problem in~\eqref{eq:gevd-multiple-outputs} is afterward recomputed given the selected sensors from~\eqref{eq:sdr}.

\subsection{Iterative reweighting and algorithm}
\label{sec:iterative-reweighting-and-algo}
\noindent
Similarly to~\cite{mehanna2013joint,hamza2019hybrid}, the all-ones matrix $\vec{1}_C\transpose{\vec{1}}_C$ in~\eqref{eq:sdr} can be replaced with a reweighting matrix $\mat{B}^{(i)} \in \R^{C \times C}$ to implement iteratively reweighted $\ell_1$-norm regularization~\cite{candes2008enhancing}. The optimization problem in~\eqref{eq:sdr} can then be iteratively solved by updating $\mat{B}^{(i)}$ as:
\begin{equation}
    B^{(i+1)}_{c_1c_2} = \frac{1}{U^{(i)}_{c_1c_2}+\epsilon}.
\end{equation}
This iteratively reweighted $\ell_1$-norm regularization procedure compensates for the inherent magnitude-dependency of the $\ell_1$-norm. Using the $\ell_1$-norm as a proxy for the $\ell_0$-norm introduces a too large penalty on the elements that have a large magnitude, while it is only relevant to know whether an element is equal to zero or not~\cite{candes2008enhancing}. The parameter $\epsilon$ avoids division by zero and is set to $10\%$ of the standard deviation of the elements of $\mat{U}$ without sensor selection (as suggested in~\cite{candes2008enhancing} and which can be easily computed using the GEVD in~\eqref{eq:gevd-multiple-outputs}). Initially, $\mat{B}^{(1)}$ is set to $\vec{1}_C\transpose{\vec{1}}_C$, i.e.,~\eqref{eq:sdr} is solved. This iterative reweighting procedure generally converges after a few iterations.

To find the optimal set of a specific number $M$ of sensors, a binary search on the hyperparameter $\mu$ of~\eqref{eq:sdr} can be performed. Once the optimal set of sensors is found, the corresponding spatio-temporal filters $\mat{W}$ can be computed by taking the $K$ GEVcs corresponding to the $K$ largest GEVls of the GEVD in~\eqref{eq:gevd-multiple-outputs}, using the reduced covariance matrices $\mat{R}^{(\textnormal{red})}_{1,2} \in \R^{ML \times ML}$, i.e., by selecting the rows and columns corresponding to the selected sensors. The complete algorithm, which is referred to as \mbox{GS-$\ell_{1,\infty}$} (GS for group-sparse) in the remainder of the paper, is summarized in Algorithm~\ref{algo:group-sparse-sensor-selection}\footnote{An open-source toolbox with the MATLAB implementation of this group-sparse sensor selection algorithm can be found online on \url{https://github.com/AlexanderBertrandLab/gsl1infSensorSelection}.}. The convex optimization problem in~\eqref{eq:sdr} is solved using the CVX toolbox~\cite{grant2014cvx,grant2008graph} and MOSEK solver~\cite{mosek2019}. 

\textbf{Remark:} It is noted that this algorithm can be easily extended to complex filter coefficients (as often found in beamforming), as the objective function of~\eqref{eq:sdr} (with the transpose replaced by Hermitian transpose) is a real-valued function, even though it is a function of complex variables, while the inequality constraints are also real. This is due to the use of the trace operator in combination with Hermitian (conjugate symmetric) complex-valued matrices.

\begin{algorithm}
		\caption{Group-sparse sensor selection for GEVD (\mbox{GS-$\ell_{1,\infty}$})}
		\label{algo:group-sparse-sensor-selection}
		\KwIn{\begin{itemize}
		    \item $\mat{R}_1,\mat{R}_2\in \R^{CL \times CL}$: to-be-discriminated covariance matrices
		    \item $M$: number of sensors to be selected
		    \item $K$: number of filters/GEVcs to take into account
		    \item $\mu_{\textnormal{LB}}, \mu_{\textnormal{UB}}$: lower and upper bounds of the binary search
		    \item $i_{\text{max}}$: maximal number of reweighting iterations
		  %  \item $i_{\text{max}}^{\text{bs}}$: maximal number of iterations in the binary search
		\end{itemize}}
		\KwOut{Optimal subset of $M$ sensors and corresponding filters/GEVCs $\mat{w} \in \R^{ML \times K}$}
		\begin{algorithmic}[1]
			\STATE Define $\epsilon$ as $10\%$ of the standard deviation of the elements of $\mat{U}$ corresponding to the solution with all sensors (as can be computed from the GEVD in~\eqref{eq:gevd-multiple-outputs}) and the tolerance $\tau$ as $10\%$ of the minimum across the diagonal of $\mat{U}$ corresponding to the solution with all sensors
			\WHILE{Not $M$ sensors selected}
			    \STATE Initialize $\mat{B}^{(1)} = \vec{1}_C\transpose{\vec{1}}_C$
    			\STATE $\mu = \mu_{\textnormal{LB}}+\frac{\mu_{\textnormal{UB}}-\mu_{\textnormal{LB}}}{2}$
    			\WHILE{\mat{U} changes and $i \leq i_{\text{max}}$}
        			\STATE Solve 
                \begin{mini*}|s|
                    {\substack{\mat{V} \in \R^{CLK\times CLK}, \mat{U} \in \R^{C \times C}}}{\trace{\left(\mat{I}_K\kron\mat{R}_2\right)\mat{V}} + \mu\trace{\mat{B}^{(i)}\mat{U}}}{}{}
                    \addConstraint{\trace{\mat{R}_1\mat{S}_{k'}\mat{V}\transpose{\mat{S}}_k}= \delta_{kk'},}{}{\quad\forall k,k' \in \{1,\dots,K\}} 
                    \addConstraint{\mat{U}\geq |\mat{S}_{k,l}\mat{V}\transpose{\mat{S}}_{k',l'}|,}{}{\quad\forall k,k' \in \{1,\dots,K\} \text{ and }\forall l,l' \in \{1,\dots,L\}}
                    \addConstraint{\mat{V} \succcurlyeq 0}
                \end{mini*}
                \STATE Update counter $i$
                \STATE Update $\mat{B}^{(i+1)}$ as:
                    \[
                        B^{(i+1)}_{c_1c_2} = \frac{1}{U^{(i)}_{c_1c_2}+\epsilon}
                    \]
    		    \ENDWHILE
    			\algstore{algo}
    		\end{algorithmic}
\end{algorithm}

\begin{algorithm}                     
    \begin{algorithmic}[1]                  
    \algrestore{algo}
    		\STATE Determine the (number of) sensors $\hat{M}$ selected by comparing the diagonal of $\mat{U}$ with the tolerance $\tau$:
    		\FOR{$c = 1$ to $C$}
    % 		\IF{$U_{cc} > \tau$} 
    		\STATE \textbf{if} $U_{cc} > \tau$ \textbf{then} $c$\textsuperscript{th} sensor selected
    % 		\ELSIF{$U_{cc} < \tau$}    
    		    \STATE \textbf{else if} $U_{cc} < \tau$ \textbf{then} $c$\textsuperscript{th} sensor eliminated
    % 		\ENDIF 
    		\ENDFOR
    		\STATE Update the regularization parameter bounds as:
    % 		\IF{$\hat{M} > M$}
    		    \STATE \textbf{if} $\hat{M} > M$ \textbf{then} $\mu_{\textnormal{LB}} = \mu$
    % 		\ELSIF{$\hat{M} < M$}    
    		    \STATE \textbf{else if} $\hat{M} < M$ \textbf{then} $\mu_{\textnormal{UB}} = \mu$
    % 		\ENDIF
			\ENDWHILE
			\STATE Compute the optimal filters as the $K$ GEVcs corresponding to the $K$ largest GEVls of the GEVD problem with reduced covariance matrices $\mat{R}^{(\textnormal{red})}_{1,2} \in \R^{ML \times ML}$:
			\[
			\mat{R}^{(\textnormal{red})}_1\mat{W} = \mat{R}^{(\textnormal{red})}_2\mat{W}\mat{\Lambda}
			\]
		\end{algorithmic}
\end{algorithm}

\subsection{Special case I: MISO filtering}
\label{sec:special-case-one-filter}
\noindent
When taking only one filter into account for the sensor selection (i.e., $K = 1$; MISO filtering), the SDR problem in~\eqref{eq:sdr} becomes:
\begin{mini}|s|
{\substack{\mat{V} \in \R^{CL\times CL}, \\ \mat{U} \in \R^{C \times C}}}{\trace{\mat{R}_2\mat{V}} + \mu\trace{\vec{1}_C\transpose{\vec{1}}_C\mat{U}}}{\label{eq:K1-sdr}}{}
\addConstraint{\trace{\mat{R}_1\mat{V}} = 1}
\addConstraint{\mat{U} \geq |\mat{S}_l\mat{V}\transpose{\mat{S}}_{l'}|, \forall l,l' \in \{1,\dots,L\}}
\addConstraint{\mat{V} \succcurlyeq 0,}
\end{mini}
with the selector-matrix $\mat{S}_l \in \R^{C \times CL}$ selecting all sensor coefficients corresponding to the $l$\textsuperscript{th} lag. This simplified problem is very similar to the approach proposed in~\cite{hamza2020sparse}, which was independently published during the consolidation of this work. However, the algorithm derived in~\cite{hamza2020sparse} has a subtle - yet crucial - difference with~\eqref{eq:K1-sdr} in the inequality constraint $\mat{U} \geq |\mat{S}_l\mat{V}\transpose{\mat{S}}_{l'}|, \forall l,l' \in \{1,\dots,L\}$. In~\cite{hamza2020sparse}, a different inequality was proposed, which only takes the diagonal elements of the different blocks of $\mat{V}$ into account, i.e., $\mat{U} \geq |\mat{S}_l\mat{V}\transpose{\mat{S}}_l|, \forall l \in \{1,\dots,L\}$, while we also take the off-diagonal elements of each block of $\mat{V}$ into account (remember: the blocks of $\mat{V}$ correspond to sensors when $K = 1$, the elements per block to different combinations of lags (see~\eqref{eq:w-definition})). While leading to fewer inequality constraints and thus resulting in a decreased computational complexity, this relaxation in~\cite{hamza2020sparse} alters the solution and leads to a suboptimal sensor selection (as empirically shown in Section~\ref{sec:benchmark-study}). The reason is that the off-diagonal blocks also appear in the first constraint of~\eqref{eq:K1-sdr}, resulting in a mismatch between both constraints. In the remainder of the paper, the variant of~\cite{hamza2020sparse} is dubbed `\mbox{GS-$\ell_{1,\infty}$-\cite{hamza2020sparse}}'.

\subsection{Special case II: purely spatial filtering}
\label{sec:special-case-two-spatial}
\noindent
In case we do not only constrain to MISO filtering ($K = 1$), but also restrict $\vec{w} \in \R^C$ to a purely spatial filter (i.e., $L = 1$),~\eqref{eq:K1-sdr} is reduced to:
\begin{mini}|s|
{\substack{\mat{V} \in \R^{C\times C}, \mat{U} \in \R^{C \times C}}}{\trace{\mat{R}_2\mat{V}} + \mu\trace{\vec{1}_C\transpose{\vec{1}}_C\mat{U}}}{\label{eq:K1L1-sdr}}{}
\addConstraint{\trace{\mat{R}_1\mat{V}} = 1}
\addConstraint{\mat{U} \geq |\mat{V}|}
\addConstraint{\mat{V} \succcurlyeq 0,}
\end{mini}
which is equivalent to the approach proposed in~\cite{hamza2019hybrid}.

\subsection{Computational complexity}
\label{sec:comp-complex-gsl1inf}
\noindent
The computational complexity of the proposed method can be computed from the complexity of interior-point method solvers for quadratic problems with quadratic constraints that are relaxed using semidefinite relaxation. In general, such problems with $N^2$ variables and $T$ quadratic constraints
can be solved to an arbitrary small accuracy $\epsilon$ with a complexity of  $\bigO\left(\text{max}(N, T)^4 N^{0.5} \log(\frac{1}{\epsilon})\right)$~\cite{luo2010semidefinite}. This leads to a complexity of $\bigO\left((CLK)^{4.5} \log(\frac{1}{\epsilon})\right)$ for our proposed \mbox{GS-$\ell_{1,\infty}$} algorithm (Algorithm~\ref{algo:group-sparse-sensor-selection}).

% *** BENCHMARK STUDY ***
\section{Benchmark study}
\label{sec:benchmark-study}
\noindent
We compare the proposed \mbox{GS-$\ell_{1,\infty}$} method with other benchmark sensor selection methods on simulated sensor data with known ground-truth\footnote{We provide an open-source MATLAB implementation of the benchmark study online on \url{https://github.com/AlexanderBertrandLab/benchmarkStudySensorSelection}.}. We use the value of the GRQ~\eqref{eq:rayleigh-quotient-multiple-outputs} (in \SI{}{\decibel}) as the performance metric (higher is better).

Besides the exhaustive search, a random search, and the \mbox{GS-$\ell_{1,\infty}$-\cite{hamza2020sparse}} method, the proposed \mbox{GS-$\ell_{1,\infty}$} method is compared with three other sensor selection methods, which are introduced in Section~\ref{sec:benchmark-sensor-selection-methods}. For the random search, the final GRQ is the mean over $1000$ random selections of sensors for a given problem. The setup of the benchmark study is described in Section~\ref{sec:setup-simulations}. The aforementioned methods are compared using only one filter (MISO filtering) in Section~\ref{sec:comp-one-filter} and using multiple filters (MIMO filtering) in Section~\ref{sec:comp-multiple-filters} (for those methods that allow for $K>1$). Finally, we provide a more in-depth comparison of the two best-performing algorithms, namely the proposed \mbox{GS-$\ell_{1,\infty}$} method and the backward greedy elimination method (see Section~\ref{sec:backward-elimination}) in Section~\ref{sec:comp-fail-case-BE}.

\subsection{Benchmark sensor selection methods}
\label{sec:benchmark-sensor-selection-methods}
\noindent
In this section, we briefly introduce other algorithms for sensor selection that will be included in the benchmark study.

\subsubsection{Greedy sensor selection methods}
\label{sec:greedy}
\noindent
Greedy sensor selection methods - also dubbed `wrapper' methods~\cite{alotaiby2015review} - sequentially select or eliminate those sensors that maximally increase or minimally decrease the objective, respectively. While these greedy approaches are computationally more efficient than the method proposed in Section~\ref{sec:optimal-selection}, due to their sequential nature, the greedy mechanism can result in suboptimal selections, as they are stuck with the selected or eliminated sensors from previous steps. The computational complexity of these greedy methods is dominated by the GEVD computation performed at each iteration, which is  $\bigO\left((CLK)^3\right)$~\cite{golub2000eigenvalue}. The greedy selection can be applied in two directions (forward or backward):

\paragraph{Forward selection (FS)}
\label{sec:forward-selection}
\noindent
The FS method starts from an empty set of sensors and sequentially adds the sensor (i.e., group of $KL$ variables) that maximally increases the objective~\eqref{eq:rayleigh-quotient-multiple-outputs}. New sensors are added until $M$ out of $C$ sensors are selected.

\paragraph{Backward elimination (BE)}
\label{sec:backward-elimination}
\noindent
The BE method starts from the full set of sensors and sequentially removes the sensor that minimally decreases the objective (i.e., the objective in~\eqref{eq:rayleigh-quotient-multiple-outputs}) until $M$ out of $C$ sensors are selected. Many variations on the FS and BE method exist, mostly presented in the context of feature selection for classification~\cite{dash1997feature}.

\subsubsection{The STECS method}
\label{sec:stecs}
\noindent
We also compare with the spatio-temporal-filtering-based channel selection (STECS) approach proposed for the GEVD problem in~\cite{qi2021spatiotemporal}. In the STECS method, initially proposed for $K=1$, the following optimization problem is solved as a regularized proxy for~\eqref{eq:rayleigh-quotient}:
\begin{mini}|s|
{\substack{\vec{w} \in \R^{CL}}}{\transpose{\vec{w}}\mat{R}_2\vec{w}+\frac{1}{\transpose{\vec{w}}\mat{R}_1\vec{w}}+\mu \norm{\vec{w}}_{1,2},}{\label{eq:stecs-optimization-problem}}{}
\end{mini}
with the $\ell_{1,2}$-norm, defined as $\norm{\vec{w}}_{1,2} = \sumlim{c = 1}{C}  \;\norm{\vec{w}_{c}}_{2}$, enforcing the group sparsity over different lags. The GRQ in~\eqref{eq:rayleigh-quotient} is split into the first two terms of~\eqref{eq:stecs-optimization-problem} as it removes the scale-invariance of $\vec{w}$, while yielding an equivalent solution~\cite{qi2021spatiotemporal}. 

The $c$-th sensor is then selected if $\norm{\vec{w}_{c}}_{2}$ is larger than a predefined tolerance $\tau$, which is set as $10\%$ of the minimum $\underset{c \in \{1,\dots,C\}}{\textnormal{min}}\norm{\vec{w}_{c}}_{2}$ of the solution with all sensors. The optimization problem in~\eqref{eq:stecs-optimization-problem} is solved using the line-search method proposed in~\cite{qi2021spatiotemporal}. The same settings as in~\cite{qi2021spatiotemporal} have been used. Similarly to the proposed method in Section~\ref{sec:iterative-reweighting-and-algo}, we use a binary search method on the regularization parameter $\mu$ to obtain the correct number of sensors $M$. Furthermore, the authors propose to use the selected sensors with~\eqref{eq:stecs-optimization-problem} for the first filter $\vec{w} \in \R^{CL}$ to recompute the solution of $\mat{W} \in \R^{CL \times K}$ for multiple filters with~\eqref{eq:gevd-multiple-outputs}, which means the selection does not take the full objective~\eqref{eq:rayleigh-quotient-multiple-outputs} into account. Lastly, note that the optimization problem~\eqref{eq:stecs-optimization-problem} is non-convex, resulting in potential convergence to non-optimal local minima.

Other sensor selection methods for GEVD problems (and in particular in biomedical applications for CSP problems) have been proposed as well~\cite{alotaiby2015review}, for example, based on filter coefficients magnitude (e.g.,~\cite{meng2009automated}), or other variants of $\ell_1$-norm regularization (e.g.,~\cite{hamza2020sparseConf,hamza2021sparse,arvaneh2011optimizing,onaran2013sparse}). However, we do not further consider these methods, as they are not designed for group-sparsity and/or the MIMO case, or have been shown to be outperformed by at least one of the aforementioned methods~\cite{qi2021spatiotemporal}.

\subsection{Setup}
\label{sec:setup-simulations}
\noindent

\subsubsection{Simulation model}
\label{sec:simulation-model}
\begin{figure}
    \centering
    \includegraphics[width=0.4\linewidth]{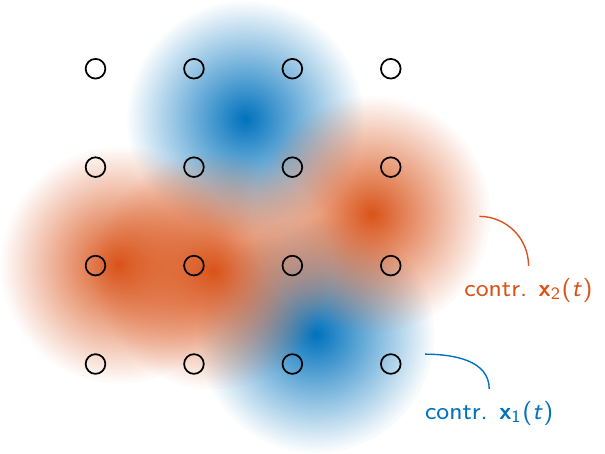}
    \caption{An exemplary generated problem with $C = 16 = 4 \times 4$ sensors, $N_1 = 2$ random signals contributing to $\vec{x}_1(t)$, and $N_2 = 3$ random signals contributing to $\vec{x}_2(t)$. Each sensor measures a mixture of the underlying sources. The brightness of the color represents the intensity of the signal as perceived by a sensor.}
    \label{fig:example-simulation}
\end{figure}

\noindent
We assume a $\sqrt{C} \times \sqrt{C}$ square grid of $C$ sensors, each of which are measuring a mixture of $N_1$ source signals to be maximized (i.e., contributing to $\vec{x}_1(t)$ and the numerator of~\eqref{eq:rayleigh-quotient-multiple-outputs}) and $N_2$ source signals to be minimized (i.e., contributing to $\vec{x}_2(t)$ and the denominator of~\eqref{eq:rayleigh-quotient-multiple-outputs}) as well as independent sensor noise. This simulated problem resembles point-source models as, for example, found in sensor networks, microphone arrays, neural activity (EEG), and telecommunications. The source signals contributing to $\vec{x}_2(t)$ have a power that is approximately 150 times larger than the source signals contributing to $\vec{x}_1(t)$. An example is given in Figure~\ref{fig:example-simulation}. In the max-SNR filtering case, one could think of the source signals contributing to $\vec{x}_1(t)$ as target signals and source signals contributing to $\vec{x}_2(t)$ as noise signals. In that case, the GRQ can be interpreted as an SNR.

Each source signal is a bandpass-filtered white Gaussian signal in a random frequency band between $1$ and $\SI{9}{\hertz}$, sampled at $\SI{20}{\hertz}$. It originates from a random location within the grid of sensors (drawn from a uniform distribution over the entire area) and propagates with an exponentially decaying amplitude to the sensors. The spread of the exponential decay is set such that the maximal attenuation is equal to a predefined attenuation of $0.5\%$. Furthermore, a source signal is measured at each sensor with a time delay linear to the distance to that source signal, such that the maximal delay is $\SI{100}{\milli\second}$ (i.e., $2$ samples). The sensor noise at each sensor is white Gaussian noise with twice the maximal attenuation as amplitude.

\subsubsection{Monte Carlo runs}
\label{sec:monte-carlo-runs}
\noindent
For each experiment, i.e., for a given number of sensors $C$, number of lags $L$, and number of filters $K$, 250 (for the $K=1$ case) and 100 (for the $K>1$ case) of the random problems in Section~\ref{sec:simulation-model} are generated, and the results for each evaluated sensor selection method are averaged across these different problems. For each of these Monte Carlo runs, unless specified otherwise, the number of signals $N_1$ and $N_2$ is randomized between $1$ and $2C$. 

\subsubsection{Hyperparameter choice}
\label{sec:hyperparam-choice}
\noindent
Table~\ref{tab:hyperparam-choice} shows the chosen hyperparameters for the different optimization-based sensor selection methods. The binary search for the \mbox{GS-$\ell_{1,\infty}$} (Algorithm~\ref{algo:group-sparse-sensor-selection}), the \mbox{GS-$\ell_{1,\infty}$-\cite{hamza2020sparse}}, and the STECS method is aborted if no solution was found after a certain number of iterations. For STECS, this number is taken much larger, which is possible due to its computational efficiency. However, this early stopping criterion leads to a limited number of cases where no solution is found for a certain $M$. To still produce a meaningful solution in those cases, a random extra sensor is added to the solution obtained for $M-1$ sensors, and the corresponding output GRQ is computed. However, when no solution is found for the lowest value of $M$ and the previous solution to still produce a meaningful solution correspondingly fails (because it relies on the solution of the lowest $M$), the results for all methods for those $M$ for which there is no solution in that particular run are removed.

Furthermore, the hyperparameter $\mu$ for the \mbox{GS-$\ell_{1,\infty}$} (Algorithm~\ref{algo:group-sparse-sensor-selection}) and \mbox{GS-$\ell_{1,\infty}$-\cite{hamza2020sparse}} algorithm is defined relative to the first target objective part of~\eqref{eq:K1-sdr} (i.e., $\trace{\mat{R}_2\mat{V}}$) for the solution with all sensors.

\begin{table}
	\centering
	\begin{tabular}{lcccc}
		\toprule
		& $\boldsymbol{\mu}_{\textnormal{\textbf{LB}}}$ &
		$\boldsymbol{\mu}_{\textnormal{\textbf{UB}}}$  & $\boldsymbol{i}_{\textnormal{\textbf{max}}}$ &  \textbf{max. it. binary search}\\
		\midrule
		\textbf{\mbox{GS-$\ell_{1,\infty}$}} & $10^{-5}$ & $100$ &  $15$ & $20$ \\
		\textbf{\mbox{GS-$\ell_{1,\infty}$-\cite{hamza2020sparse}}} & $10^{-5}$ & $10^4$ & $15$ & $20$ \\
		\textbf{STECS} & $0$ & $10^{16}$ & / & $200$ \\
		\hline
	\end{tabular}
	\caption{The chosen hyperparameters in the different optimization-based sensor selection methods.}
	\label{tab:hyperparam-choice}
\end{table}

\subsubsection{Statistical comparison}
\label{sec:statistical-comparison}
\noindent
To identify statistically significant differences based on hypothesis testing, we use a linear mixed-effect model (LMEM)~\cite{galecki2013linear}. Such an LMEM allows the exploitation of all structure in the data by modeling the obtained GRQ as a function of the method while taking the variation due to the different Monte Carlo runs and the effect of a different number of selected sensors $M$ into account as random factors. The following LMEM is chosen based on the Akaike information criterion (AIC), which takes the model fit and complexity into account:
\[
\textnormal{GRQ} \sim 1+\textnormal{method} + (M|\textnormal{run}).
\]
This notation is often used in LMEMs to reflect that the GRQ is modeled with the method as a fixed effect, the number of selected sensors $M$ as random slope (i.e., the GRQ can vary as a function of $M$ independently for each method), and the run as a random intercept. Per fixed term, the estimated regression coefficients ($\beta$), standard errors (SE), degrees of freedom (DF), t-value, and p-value are reported. If a significant effect between the different methods is found, we use an additional Tukey-adjusted post-hoc test to assess the pairwise differences between individual methods. The significance level is set to $\alpha = 0.05$. All statistical analyses are performed using the R software package and the \texttt{nlme} and \texttt{emmeans} packages.

In the statistical hypothesis testing, we limit the number of selected sensors to $\frac{C}{2}$, as we consider this lower half range much more relevant in the context of sensor selection than the upper half range. Typically, one wants to drastically reduce the number of required sensors, i.e., below half of the number of available sensors.

\subsection{Comparison in the MISO case ($K = 1$)}
\label{sec:comp-one-filter}
\begin{figure}
    \centering
    \includegraphics[width=0.7\linewidth]{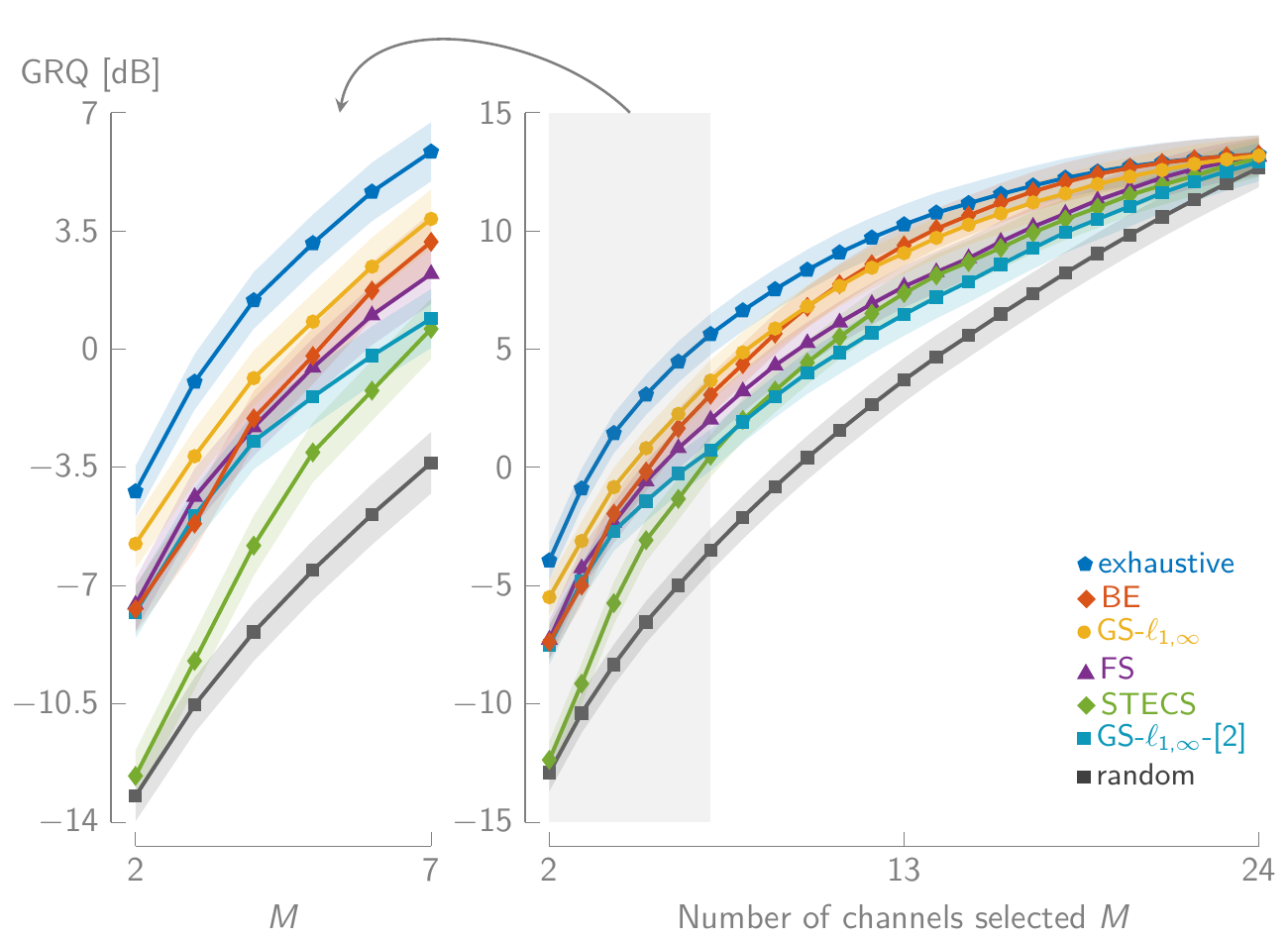}
    \caption{The output GRQ (mean over 250 runs) as a function of $M$ for the different sensor selection methods when $C = 25, L = 3, K = 1$. The shading represents the standard error on the mean.}
    \label{fig:results-c25-l3-k1}
\end{figure}

\noindent
First, we evaluate and compare the presented methods in the first special MISO case of Section~\ref{sec:special-case-one-filter}, where only one filter (first GEVc) is taken into account, i.e., $K = 1$. In this case, we can also include the comparison with \mbox{GS-$\ell_{1,\infty}$-\cite{hamza2020sparse}}, which was designed specifically for this case. We choose $C = 25,L = 3$ and look for the optimal sensor selection for $M$ ranging from $2$ to $24$. Figure~\ref{fig:results-c25-l3-k1} shows the output GRQs (in \SI{}{\decibel}) as a function of $M$ for each separate method (mean over 250 Monte Carlo runs $\pm$ the standard error on the mean). Table~\ref{tab:stat-test-c25-l3-k1} shows the outcome of the statistical analysis, for $M$ ranging from $2$ to $12$ (see Section~\ref{sec:statistical-comparison}). All presented methods achieve significantly higher (better) GRQ than random selection but lower (worse) GRQ than the optimal solution obtained through an exhaustive search over all possible combinations (Table~\ref{tab:pairwise-comp-c25-l3-k1}).

\begin{table}
    \begin{subtable}[h]{1\linewidth}
        \centering
        % \begin{adjustbox}{width=1\textwidth}
        \begin{tabular}{lccccc}
    		\toprule
    		\textbf{Fixed-effect term} &
    		$\boldsymbol{\beta}$ & \textbf{SE} & \textbf{DF} &  \textbf{t-value} & \textbf{p-value}\\
    		\midrule
    		intercept & $-5.38$ & $0.79$ & $18729$ & $-6.80$ & $< 0.0001$\\
    		method = exhaustive$-$BE & $2.51$ & $0.07$ & \multirow{6}{*}{\rule{0.5pt}{114pt}} & $34.94$ & $< 0.0001$\\
    		method = exhaustive$-$FS & $3.35$ & \multirow{5}{*}{\rule{0.5pt}{95pt}} & & $46.71$ & $< 0.0001$\\
    		method = exhaustive$-$STECS & $5.48$ & & & $76.30$ & $< 0.0001$\\
    		method = exhaustive$-$random & $8.73$ & & & $121.59$ & $< 0.0001$\\
    		method = exhaustive$-$\mbox{GS-$\ell_{1,\infty}$-\cite{hamza2020sparse}} & $4.33$ & & & $60.36$ & $< 0.0001$\\
            method = exhaustive$-$\mbox{GS-$\ell_{1,\infty}$} & $1.83$ & & & $25.46$ & $< 0.0001$\\
    		\hline
    	\end{tabular}
    % 	\end{adjustbox}
    	\caption{}
    	\label{tab:lmem-c25-l3-k1}
    \end{subtable}
    
    \begin{subtable}[h]{1\linewidth}
        \centering
        % \begin{adjustbox}{width=1\textwidth}
        \begin{tabular}{lccccccc}
    		\toprule
    		& \textbf{exhaustive} & \textbf{BE} & \textbf{FS} & \textbf{STECS} &  \textbf{random} & \textbf{\mbox{GS-$\ell_{1,\infty}$-\cite{hamza2020sparse}}} & \textbf{\mbox{GS-$\ell_{1,\infty}$}}\\
    		\midrule
    		\textbf{exhaustive} & / &  {\color{g}$2.51/*$} & {\color{g}$3.53/*$} & {\color{g}$5.48/*$} & {\color{g}$8.73/*$} & {\color{g}$4.33/*$} & {\color{g}$1.83/*$} \\
    		\textbf{BE} & {\color{r}$-2.51/*$} & / & {\color{g}$0.85/*$} & {\color{g}$2.97/*$} & {\color{g}$6.22/*$} & {\color{g}$1.83/*$} & ${\color{r}-0.68/*}$ \\
    		\textbf{FS} & {\color{r}$-3.53/*$} & {\color{r}$-0.85/*$} & / & {\color{g}$2.13/*$} & {\color{g}$5.38/*$} & {\color{g}$0.98/*$} & {\color{r}$-1.53/*$} \\
    		\textbf{STECS} & {\color{r}$-5.48/*$} & {\color{r}$-2.97/*$} & {\color{r}$-2.13/*$} & / & {\color{g}$3.25/*$} & {\color{r}$-1.15/*$} & {\color{r}$-3.65/*$} \\
    		\textbf{random} & {\color{r}$-8.73/*$} & {\color{r}$-6.22/*$} & {\color{r}$-5.38/*$} & {\color{r}$-3.25/*$} & / & {\color{r}$-4.40/*$} & {\color{r}$-6.90/*$}\\
    		\textbf{\mbox{GS-$\ell_{1,\infty}$-\cite{hamza2020sparse}}} & {\color{r}$-4.33/*$} & {\color{r}$-1.83/*$} & {\color{r}$-0.98/*$} & {\color{g}$1.15/*$} & {\color{g}$4.40/*$} & / & {\color{r}$-2.51/*$}\\
    		\textbf{\mbox{GS-$\ell_{1,\infty}$}} & {\color{r}$-1.83/*$} & {\color{g}$0.68/*$} & {\color{g}$1.53/*$} & {\color{g}$3.65/*$} & {\color{g}$6.90/*$} & {\color{g}$2.51/*$} & / \\
    		\hline
    	\end{tabular}
    % 	\end{adjustbox}
    	\caption{}
    	\label{tab:pairwise-comp-c25-l3-k1}
     \end{subtable}
     \caption{(a) The LMEM fixed-effect outcomes for $M = 2$ to $12$ when $C = 25, L = 3, K = 1$. (b) The pairwise differences, showing the estimated difference between average GRQ  (method in row $-$ method in column)/p-value per pair of methods (p-values $< 0.0001$ are indicated with $*$). Statistically significant differences are color coded. Values in {\color{g}green}/{\color{r}red} indicate that the method in the row outperforms/is outperformed by the method in the column.}
     \label{tab:stat-test-c25-l3-k1}
\end{table}

The greedy sensor selection methods suffer from intrinsic limitations, i.e., they depend on previous choices in their sequential procedure. For example, the FS method starts with a GRQ close to optimal but diverges from the optimal exhaustive solution when $M$ increases, and the other way around for the BE method. However, the FS method achieves overall lower GRQ than the BE method, which is also confirmed by the statistical testing in Table~\ref{tab:pairwise-comp-c25-l3-k1}. This could be due to the fact the FS method is limited to selecting one sensor at a time, which hampers its capacity to probe combined effects of multiple sensors.

Furthermore, Figure~\ref{fig:results-c25-l3-k1} and Table~\ref{tab:pairwise-comp-c25-l3-k1} show that the \mbox{\mbox{GS-$\ell_{1,\infty}$-\cite{hamza2020sparse}}} method is outperformed by all other methods, except by the STECS method. Our proposed method (significantly) outperforms \mbox{GS-$\ell_{1,\infty}$-\cite{hamza2020sparse}} across all $M$. This is an effect of dropping the off-diagonal blocks of the inequality constraints in~\eqref{eq:K1-sdr}. However, the gap between both methods becomes smaller for lower $M$ (see also Figure~\ref{fig:results-c25-l3-k1}). Similarly, the STECS method is outperformed, especially for low $M$, by all other methods, suffering from its non-convex objective function. This method achieves slightly higher GRQ than the \mbox{GS-$\ell_{1,\infty}$-\cite{hamza2020sparse}} method for most larger $M$, but it achieves much lower GRQ for small $M$. As a result, there is also a significant difference observed across all $M$ between $2$ and $12$ between the STECS and \mbox{GS-$\ell_{1,\infty}$-\cite{hamza2020sparse}} method (Table~\ref{tab:pairwise-comp-c25-l3-k1}).

A remarkable conclusion is that the greedy BE method significantly outperforms almost all other state-of-the-art methods, including \mbox{GS-$\ell_{1,\infty}$-\cite{hamza2020sparse}} and STECS, which have not been benchmarked in a group-sparse setting against BE in the corresponding original papers~\cite{hamza2020sparse} and~\cite{qi2021spatiotemporal}, respectively. The only method that significantly outperforms the BE method is our proposed \mbox{GS-$\ell_{1,\infty}$} algorithm. Interestingly, although the BE method seems to perform slightly better for larger $M$, the \mbox{GS-$\ell_{1,\infty}$} method seems to perform better than the BE method for \emph{small} $M$ especially, explaining the statistically significant difference. The gap between both methods is also larger for these small $M$ than for large $M$. From an application-based point of view, these smaller $M$ - below half of the total number of sensors $C$ - are often targeted in practice. Indeed, sensor selection is typically performed to substantially decrease the number of required sensors, not to remove only a few sensors. Although the heuristic BE method is computationally much more efficient than the optimization-based \mbox{GS-$\ell_{1,\infty}$} method, it thus performs worse than the \mbox{GS-$\ell_{1,\infty}$} method when it most matters. Lastly, it is interesting to identify in how many and in which cases a sensor selection method completely fails. For example, one could define a failure as more than $\SI{10}{\decibel}$ difference with the exhaustive method. Using this rule, the fail rate for the BE method across all runs and again for $M$ between $2$ and $12$ is $3.71\%$, while this is $0.44\%$ for the \mbox{GS-$\ell_{1,\infty}$} method. Thus, the BE method has almost 10 times more severe fail cases than the \mbox{GS-$\ell_{1,\infty}$} method. While these percentages might seem marginal at first sight, it should be taken into account that this percentage is biased by the highly randomized simulated scenarios. After a closer look, these fail cases turn out to mainly occur in cases where the covariance matrix $\mat{R}_2$ is ill-conditioned, which is not necessarily a rare case in practical settings, for example, as found in miniaturized EEG sensor networks~\cite{narayanan2020analysis}. In Section~\ref{sec:comp-fail-case-BE}, we further analyze these particular fail cases and provide a more extensive discussion.

\subsection{Comparison in the MIMO case ($K > 1$)}
\label{sec:comp-multiple-filters}
\begin{figure}
    \centering
    \includegraphics[width=0.5\linewidth]{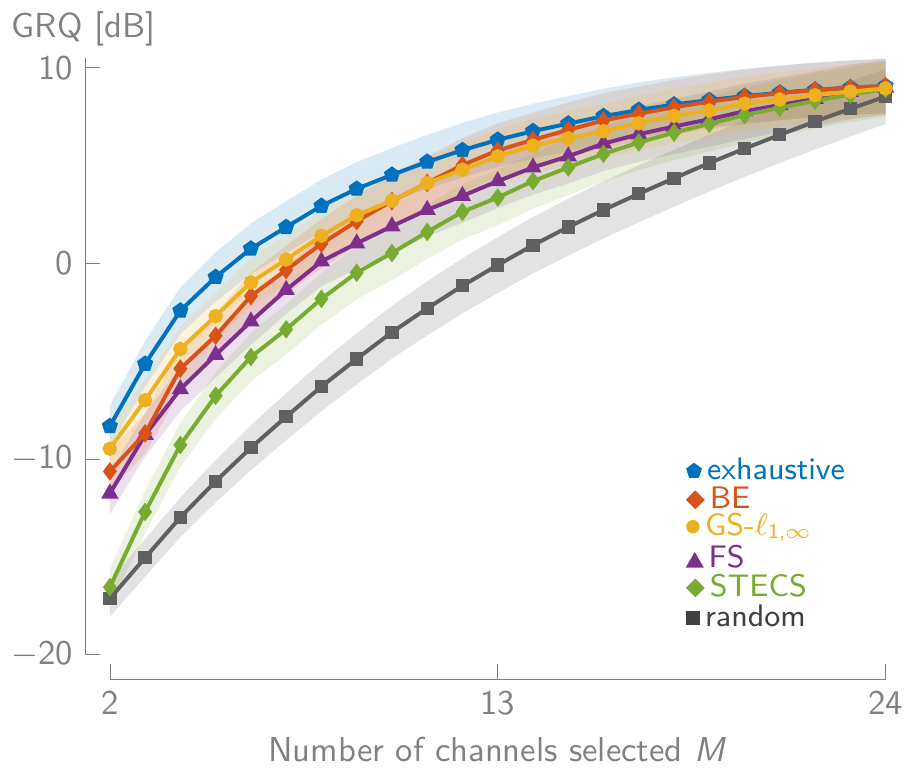}
    \caption{The output GRQ (mean over 100 runs) as a function of $M$ for the different sensor selection methods when $C = 25, L = 2, K = 2$. The shading represents the standard error on the mean.}
    \label{fig:results-c25-l2-k2}
\end{figure}

\noindent
Figure~\ref{fig:results-c25-l2-k2} and Table~\ref{tab:lmem-c25-l2-k2} show the results of 100 Monte Carlo simulations with $C = 25, L = 2$, and $K = 2$, i.e., the more general case where $K > 1$. As the \mbox{GS-$\ell_{1,\infty}$-\cite{hamza2020sparse}} method was only proposed for $K = 1$, it is not included in these simulations. 

\begin{table}
    \begin{subtable}[h]{1\linewidth}
        \centering
        % \begin{adjustbox}{width=1\textwidth}
        \begin{tabular}{lccccc}
    		\toprule
    		\textbf{Fixed-effect term} &
    		$\boldsymbol{\beta}$ & \textbf{SE} & \textbf{DF} &  \textbf{t-value} & \textbf{p-value}\\
    		\midrule
    		intercept & $-14.64$ & $0.91$ & $6393$ & $-16.01$ & $< 0.0001$\\
    		method = exhaustive$-$BE & $2.10$ & $0.11$ & \multirow{5}{*}{\rule{0.5pt}{95pt}} & $18.54$ & $< 0.0001$\\
    		method = exhaustive$-$FS & $3.18$ & \multirow{4}{*}{\rule{0.5pt}{76pt}} & & $28.04$ & $< 0.0001$\\
    		method = exhaustive$-$STECS & $5.35$ & & & $47.15$ & $< 0.0001$\\
    		method = exhaustive$-$random & $9.09$ & & & $80.17$ & $< 0.0001$\\
    		method = exhaustive$-$\mbox{GS-$\ell_{1,\infty}$} & $1.51$ & & & $13.36$ & $< 0.0001$\\
    		\hline
    	\end{tabular}
    % 	\end{adjustbox}
    	\caption{}
    	\label{tab:lmem-c25-l2-k2}
    \end{subtable}
    
    \begin{subtable}[h]{1\linewidth}
        \centering
        % \begin{adjustbox}{width=1\textwidth}
        \begin{tabular}{lcccccc}
    		\toprule
    		& \textbf{exhaustive} & \textbf{BE} & \textbf{FS} & \textbf{STECS} &  \textbf{random} & \textbf{\mbox{GS-$\ell_{1,\infty}$}} \\
    		\midrule
    		\textbf{exhaustive} & / &  {\color{g}$2.10/*$} & {\color{g}$3.18/*$} & {\color{g}$5.35/*$} & {\color{g}$9.09/*$} & {\color{g}$1.51/*$} \\
    		\textbf{BE} & {\color{r}$-2.10/*$} & / & {\color{g}$1.08/*$} & {\color{g}$3.24/*$} & {\color{g}$6.99/*$} & {\color{r}$-0.59/*$} \\
    		\textbf{FS} & {\color{r}$-3.18/*$} & {\color{r}$-1.08/*$} & / & {\color{g}$2.17/*$} & {\color{g}$5.91/*$} & {\color{r}$-1.66/*$} \\
    		\textbf{STECS} & {\color{r}$-5.35/*$} & {\color{r}$-3.24/*$} & {\color{r}$-2.17/*$} & / & {\color{g}$3.75/*$} & {\color{r}$-3.83/*$} \\
    		\textbf{random} & {\color{r}$-9.09/*$} & {\color{r}$-6.99/*$} & {\color{r}$-5.91/*$} & {\color{r}$-3.75/*$} & / & {\color{r}$-7.58/*$}\\
    		\textbf{\mbox{GS-$\ell_{1,\infty}$}} & {\color{r}$-1.51/*$} & {\color{g}$0.59/*$} & {\color{g}$1.66/*$} & {\color{g}$3.83/*$} & {\color{g}$7.58/*$} & /\\
    		\hline
    	\end{tabular}
    % 	\end{adjustbox}
    	\caption{}
    	\label{tab:pairwise-comp-c25-l2-k2}
     \end{subtable}
    \caption{(a) The LMEM fixed-effect outcomes for $M = 2$ to $12$ when $C = 25, L = 2, K = 2$ (\mbox{GS-$\ell_{1,\infty}$-\cite{hamza2020sparse}} is omitted as it is only defined for $K = 1$). (b) The pairwise differences, showing the estimated difference between average GRQ  (method in row $-$ method in column)/p-value per pair of methods (p-values $< 0.0001$ are indicated with $*$). Statistically significant differences are color coded. Values in {\color{g}green}/{\color{r}red} indicate that the method in the row outperforms/is outperformed by the method in the column.}
     \label{tab:stat-test-c25-l2-k2}
\end{table}

First of all, the results confirm that the proposed extension to MIMO filtering in Section~\ref{sec:optimal-selection} is valid, as the \mbox{GS-$\ell_{1,\infty}$} method still obtains GRQs close to those of the optimal exhaustive solution. Furthermore, the results are in line with Section~\ref{sec:comp-one-filter}. The BE and \mbox{GS-$\ell_{1,\infty}$} method again show a statistically significant difference when evaluated across $M$ between $2$ and $12$ (Table~\ref{tab:pairwise-comp-c25-l2-k2}), confirming that the latter has the advantage for the more relevant low $M$. Finally, both methods significantly outperform the other benchmark methods (except the exhaustive search).

\subsection{Comparison of \mbox{GS-$\ell_{1,\infty}$} with BE}
\label{sec:comp-fail-case-BE}
\noindent
In this section, we zoom in on the comparison between the BE and \mbox{GS-$\ell_{1,\infty}$} method, as these two methods achieve the highest GRQ in the previous simulations. As explained in Section~\ref{sec:greedy}, the BE method can suffer from its greedy sequential selection, where previously eliminated sensors can not be recovered when selecting a lower number of sensors. This inherent disadvantage of the BE method can lead to various fail cases (defined here as $>\SI{10}{\decibel}$ difference with the optimal exhaustive solution). After closer inspection, we identified that the majority of the fail cases ($73.53\%$ of all the fail cases in the previous simulations) corresponded to scenarios in which the matrix $\mat{R}_2$ was ill-conditioned, i.e., where there was a large difference between the largest and smallest eigenvalue(s).

\begin{figure}
    \centering
    \includegraphics[width=0.5\linewidth]{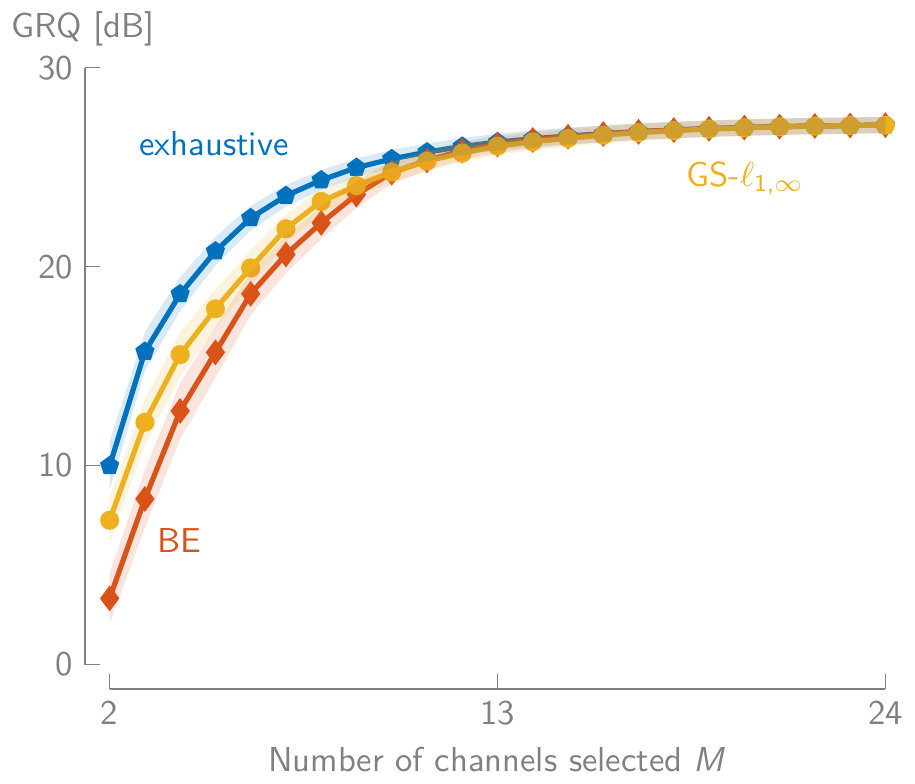}
    \caption{While the BE and \mbox{GS-$\ell_{1,\infty}$} method performs on par for large $M$, the \mbox{GS-$\ell_{1,\infty}$} method starts to outperform the BE method for smaller $M$ in the ill-conditioned $\mat{R}_2$ covariance matrix case (mean $\pm$ standard error on the mean).}
    \label{fig:results-rank-def-noise}
\end{figure}

To thoroughly test this case, we compare the BE method to the \mbox{GS-$\ell_{1,\infty}$} method on the subset of 60 simulations of Section~\ref{sec:comp-one-filter} where $2 \leq N_2 \leq 12$, i.e., where the number of signals contributing to $\vec{x}_2(t)$ is less than half of the $C = 25$ sensors. These cases correspond to ill-conditioned covariance matrices $\mat{R}_2$ in the denominator of the GRQ, where the smallest eigenvalues are determined solely by white Gaussian sensor noise (see Section~\ref{sec:simulation-model}). The results are shown in Figure~\ref{fig:results-rank-def-noise}, where also the performance of the exhaustive solution is shown as a reference. For large $M$, both methods perform similarly to the exhaustive solution, with very little change in GRQ for increasing values of $M$. When $M$ decreases, both methods start to diverge from the exhaustive solution. However, the BE method achieves lower GRQ than the \mbox{GS-$\ell_{1,\infty}$} method for smaller $M$. This is confirmed by the LMEM including only those two methods, as there is again a significant effect of the method, i.e., the \mbox{GS-$\ell_{1,\infty}$} method outperforms the BE method (Table~\ref{tab:lmem-comp-fail-case-be}).

\begin{table}
	\centering
	\begin{tabular}{lccccc}
		\toprule
		\textbf{Fixed-effect term} &
		$\boldsymbol{\beta}$ & \textbf{SE} & \textbf{DF} &  \textbf{t-value} & \textbf{p-value}\\
		\midrule
		intercept & $31.30$ & $0.44$ & $1259$ & $71.47$ & $< 0.0001$\\
		method = GS-$\ell_{1,\infty}-$BE & $1.53$ & $0.19$ & $1259$ & $8.20$ & $<0.0001$\\
		\hline
	\end{tabular}
	\caption{The LMEM outcomes when including only the BE and \mbox{GS-$\ell_{1,\infty}$} methods in the case with an ill-conditioned covariance matrix $\mat{R}_2$ in the denominator (for $M = 2$ to $12$).}
	\label{tab:lmem-comp-fail-case-be}
\end{table}

\begin{figure*}
    \centering
    \includegraphics[width=0.825\linewidth]{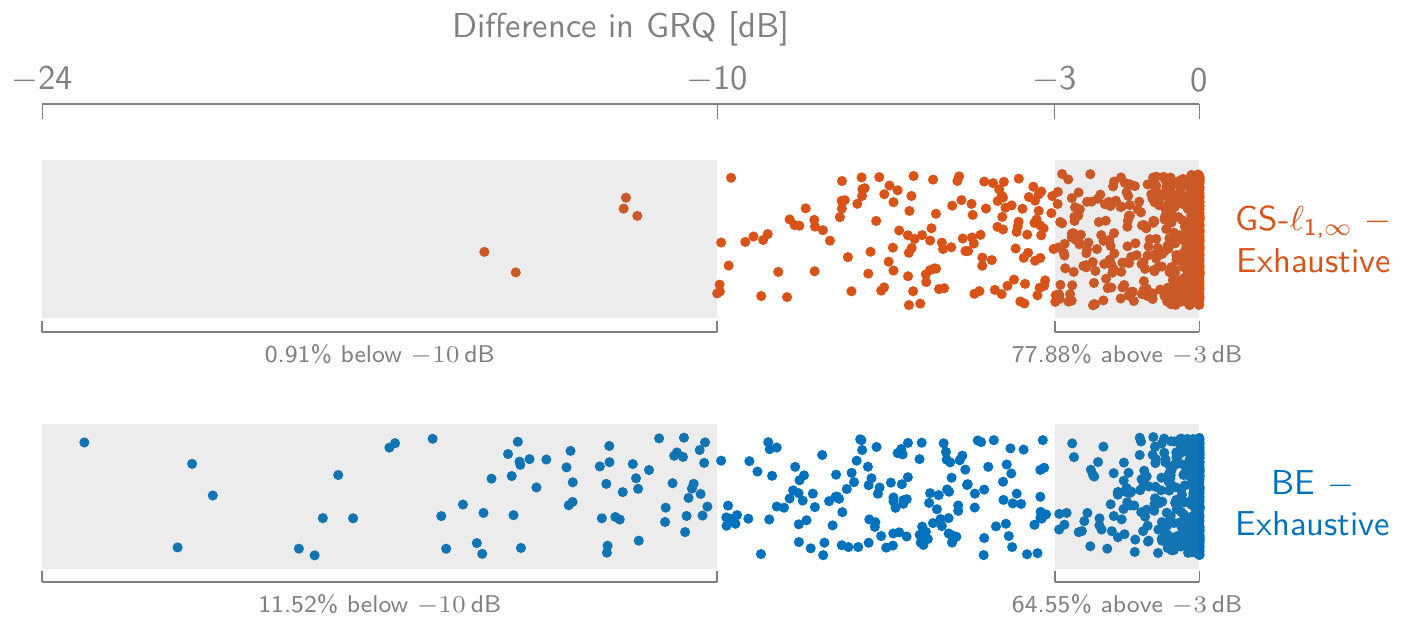}
    \caption{The BE method shows more outlying negative differences in GRQ (across all runs and $M$ between $2$ and $12$) with the exhaustive solution than the \mbox{GS-$\ell_{1,\infty}$} method when the covariance matrix in the denominator of the GRQ is ill-conditioned.}
    \label{fig:diff-snr-rank-def-noise}
\end{figure*}

Figure~\ref{fig:diff-snr-rank-def-noise} shows the differences in GRQ across all runs and $M$ between $2$ and $12$ between the BE/\mbox{GS-$\ell_{1,\infty}$} method and the exhaustive solution when $2 \leq N_2 \leq 12$. The BE method has a heavier tail with more outlying negative differences with the exhaustive solution than the \mbox{GS-$\ell_{1,\infty}$} method. Of all runs with $2 \leq N_2 \leq 12$, there is a fail rate of $11.52\%$ for the BE method, while this is only $0.91\%$ for the \mbox{GS-$\ell_{1,\infty}$} method. To summarize, when there is an ill-conditioned covariance matrix $\mat{R}_2$ in the denominator of the GRQ, the \mbox{GS-$\ell_{1,\infty}$} method is more robust than the BE method.

\section{Example of sensor selection on real-world data}
\label{sec:epilepsy}
\noindent
The benchmark study in Section~\ref{sec:benchmark-study} was performed on simulated data, which allowed us to generate a large number of simulations that are generic enough to apply to many sensor selection problems that arise in different signal processing domains. In this section, we show an example of sensor selection in the context of mobile epileptic seizure monitoring. More specifically, the task requires to design a spatio-temporal filter that amplifies multi-channel EEG data during seizures while maximally attenuating peak interferers~\cite{dan2020efficient}. The solution is found through max-SNR filtering and can be solved using the GEVD framework described in this paper. More information about the context, problem, and data can be found in~\cite{dan2020efficient}.

In the following example, we investigate the effect of the reduction of EEG channels on subject three of the study in~\cite{dan2020efficient}, aiming to design a mobile EEG setup. The data contains 16 channels (i.e., $C = 16$). Five time lags are used per channel (i.e., $L = 5$), while two output filters are computed (i.e., $K = 2$).

\begin{figure*}
    \centering
    \includegraphics[width=0.5\linewidth]{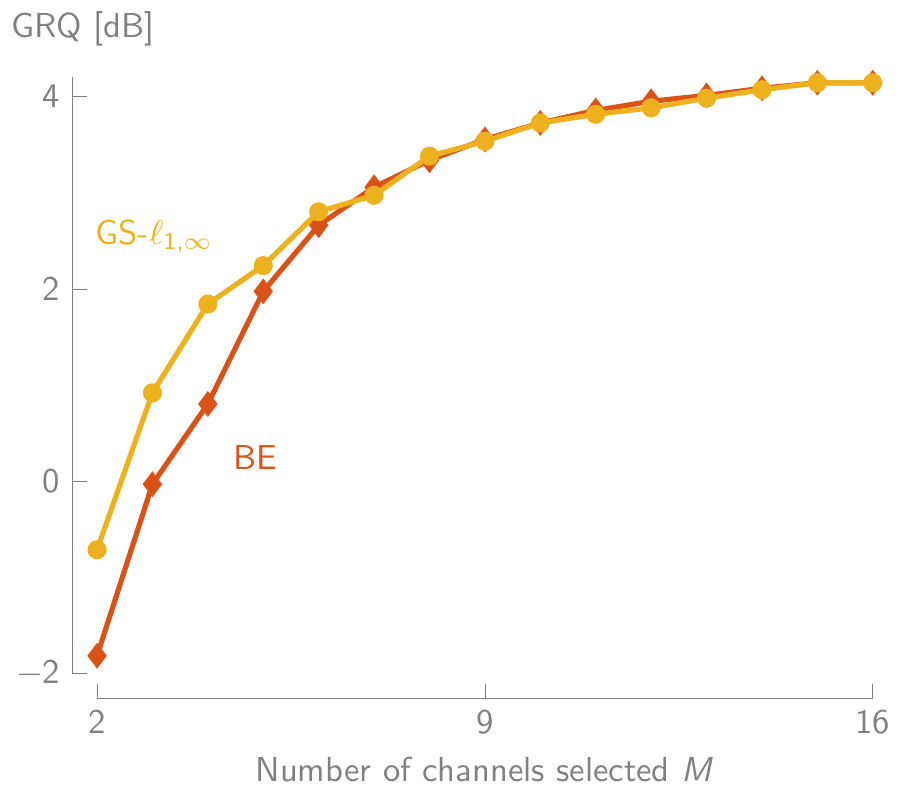}
    \caption{The \mbox{GS-$\ell_{1,\infty}$} method outperforms the BE method for a low number of channels also on an example with real-world data collected on a patient with epilepsy ($C = 16, L = 5, K = 2$).}
    \label{fig:seizure}
\end{figure*}

Figure~\ref{fig:seizure} shows the GRQ as a function of the number of selected channels for the GS-$\ell_{1,\infty}$ and BE methods, which performed best in the benchmark study (see Section~\ref{sec:comp-one-filter} to \ref{sec:comp-fail-case-BE}). The results are in line with the benchmark study. Both methods obtain similar GRQ across the whole range of selected channels, but the GS-$\ell_{1,\infty}$ method outperforms the greedy BE method for a low number of channels $M$. This confirms that the developed method performs as expected, also on sensor selection problems on real-world data.

% *** CONCLUSION ***
\section{Conclusion}
\label{sec:conclusion}
\noindent
In this paper, we proposed a group-sparse variable selection method using the $\ell_{1,\infty}$-norm for GRQ optimization and GEVD problems applied in the context of sensor selection. This group-sparsity does not only allow to extend spatial to spatio-temporal filtering but also to take multiple filters (eigenvectors) into account and thus extend MISO to MIMO filtering. The latter is essential in various other applications, such as selecting sensors across different filterbands in CSP applications~\cite{geirnaert2020fast,blankertz2007optimizing}.

We have extensively compared the proposed \mbox{GS-$\ell_{1,\infty}$} method with various other sensor selection methods (greedy, optimization-based, \dots). Remarkably, the simple greedy BE method outperformed all methods from the state of the art, except the proposed \mbox{GS-$\ell_{1,\infty}$} method. While the heuristic BE method is computationally more efficient, it performs worse than the \mbox{GS-$\ell_{1,\infty}$} method for smaller numbers of selected sensors, and with a higher probability to completely fail. We have shown that one specific fail case of the BE method is when the covariance matrix in the denominator of the GRQ is ill-conditioned. 

As the BE method is less robust than the proposed \mbox{GS-$\ell_{1,\infty}$} method, the latter is the preferred choice when performing variable selection, in particular if the number of desired variables is small compared to the total number of variables.

% *** AUTHOR CONTRIBUTIONS ***
\section*{Author contributions}
\noindent
\textbf{Jonathan Dan:} Conceptualization, Methodology, Software, Validation, Formal Analysis, Writing - Original Draft, Writing - Review \& Editing. \mbox{\textbf{Simon Geirnaert:}} Conceptualization, Methodology, Software, Validation, Formal Analysis, Writing - Original Draft, Writing - Review \& Editing. \textbf{Alexander Bertrand:} Conceptualization, Methodology, Formal Analysis, Writing - Review \& Editing, Supervision

% *** ACKNOWLEDGEMENTS ***
\section*{Acknowledgements}
\noindent
This work was supported by an Aspirant Grant from the Research Foundation - Flanders (FWO) (for S. Geirnaert - 1136219N), by VLAIO and Byteflies through a Baekeland grant (HBC.2018.0189) (for J. Dan), FWO project nr. G0A4918N, the European Research Council (ERC) under the European Union’s Horizon 2020 Research and Innovation Programme (grant agreement No 802895), and the Flemish Government (AI Research Program).

\bibliographystyle{elsarticle-num} 
\bibliography{biblio}

\begin{thebibliography}{10}
\expandafter\ifx\csname url\endcsname\relax
  \def\url#1{\texttt{#1}}\fi
\expandafter\ifx\csname urlprefix\endcsname\relax\def\urlprefix{URL }\fi
\expandafter\ifx\csname href\endcsname\relax
  \def\href#1#2{#2} \def\path#1{#1}\fi

\bibitem{chepuri2015sparsity}
S.~P. Chepuri, G.~Leus, {Sparsity-Promoting Sensor Selection for Non-Linear
  Measurement Models}, IEEE Trans. Signal Process. 63~(3) (2015) 684--698.
\newblock \href {https://doi.org/10.1109/TSP.2014.2379662}
  {\path{doi:10.1109/TSP.2014.2379662}}.

\bibitem{hamza2020sparse}
S.~A. {Hamza}, M.~G. {Amin}, {Sparse Array Beamforming Design for Wideband
  Signal Models}, IEEE Trans. Aerosp. Electron. Syst. 57~(2) (2021) 1211--1226.
\newblock \href {https://doi.org/10.1109/TAES.2020.3037409}
  {\path{doi:10.1109/TAES.2020.3037409}}.

\bibitem{gao2018sparse}
M.~Gao, K.~F.~C. Yiu, S.~Nordholm, {On the Sparse Beamformer Design}, Sensors
  18~(10) (2018).
\newblock \href {https://doi.org/10.3390/s18103536}
  {\path{doi:10.3390/s18103536}}.

\bibitem{wanlu2019controllable}
W.~Shi, Y.~Li, L.~Zhao, X.~Liu, {Controllable Sparse Antenna Array for Adaptive
  Beamforming}, IEEE Access 7 (2019) 6412--6423.
\newblock \href {https://doi.org/10.1109/ACCESS.2018.2889877}
  {\path{doi:10.1109/ACCESS.2018.2889877}}.

\bibitem{hamza2018optimum}
S.~A. Hamza, M.~G. Amin, {Optimum sparse array receive beamforming for wideband
  signal model}, in: Proc. of the 52nd ACSSC, 2018, pp. 89--93.
\newblock \href {https://doi.org/10.1109/ACSSC.2018.8645552}
  {\path{doi:10.1109/ACSSC.2018.8645552}}.

\bibitem{hamza2019sparse}
S.~A. Hamza, M.~G. Amin, {Sparse Array DFT Beamformers for Wideband Sources},
  in: Proc. of the IEEE RadarConf19, 2019, pp. 1--5.
\newblock \href {https://doi.org/10.1109/RADAR.2019.8835749}
  {\path{doi:10.1109/RADAR.2019.8835749}}.

\bibitem{hamza2020sparseConf}
S.~A. Hamza, M.~G. Amin, {Sparse Array Receiver Beamformer Design for
  Multi-Functional Antenna}, in: Proc. of EUSIPCO 2020, 2021, pp. 1836--1840.
\newblock \href {https://doi.org/10.23919/Eusipco47968.2020.9287795}
  {\path{doi:10.23919/Eusipco47968.2020.9287795}}.

\bibitem{hamza2021sparse}
S.~A. Hamza, W.~Zhai, X.~Wang, M.~G. Amin, {Sparse Array Transceiver Design for
  Enhanced Adaptive Beamforming in MIMO Radar}, in: Proc. of ICASSP 2021, 2021,
  pp. 4410--4414.
\newblock \href {https://doi.org/10.1109/ICASSP39728.2021.9414650}
  {\path{doi:10.1109/ICASSP39728.2021.9414650}}.

\bibitem{zhai2021cognitive}
W.~Zhai, X.~Wang, S.~A. Hamza, M.~G. Amin, {Cognitive-Driven Optimization of
  Sparse Array Transceiver for MIMO Radar Beamforming}, in: Proc. of the IEEE
  RadarConf21, 2021, pp. 1--6.
\newblock \href {https://doi.org/10.1109/RadarConf2147009.2021.9455310}
  {\path{doi:10.1109/RadarConf2147009.2021.9455310}}.

\bibitem{alotaiby2015review}
T.~Alotaiby, F.~E. El-Samie, S.~A. Alshebeili, I.~Ahmad, {A review of channel
  selection algorithms for EEG signal processing}, EURASIP J. Adv. Signal
  Process.~(66) (2015).
\newblock \href {https://doi.org/10.1186/s13634-015-0251-9}
  {\path{doi:10.1186/s13634-015-0251-9}}.

\bibitem{narayanan2020analysis}
A.~M. Narayanan, A.~Bertrand, {Analysis of Miniaturization Effects and Channel
  Selection Strategies for EEG Sensor Networks with Application to Auditory
  Attention Detection}, IEEE Trans. Biomed. Eng. 67~(1) (2020) 234--244.
\newblock \href {https://doi.org/10.1109/TBME.2019.2911728}
  {\path{doi:10.1109/TBME.2019.2911728}}.

\bibitem{narayanan2020optimal}
A.~M. Narayanan, P.~Patrinos, A.~Bertrand, {Optimal Versus Approximate Channel
  Selection Methods for EEG Decoding With Application to Topology-Constrained
  Neuro-Sensor Networks}, IEEE Trans. Neural Syst. Rehabilitation Eng. 29
  (2021) 92--102.
\newblock \href {https://doi.org/10.1109/TNSRE.2020.3035499}
  {\path{doi:10.1109/TNSRE.2020.3035499}}.

\bibitem{dan2020efficient}
J.~Dan, B.~Vandendriessche, W.~V. Paesschen, D.~Weckhuysen, A.~Bertrand,
  {Computationally-Efficient Algorithm for Real-Time Absence Seizure Detection
  in Wearable Electroencephalography}, Int. J. Neural Syst. 30~(11) (2020)
  2050035.
\newblock \href {https://doi.org/10.1142/S0129065720500355}
  {\path{doi:10.1142/S0129065720500355}}.

\bibitem{bertrand2011applications}
A.~{Bertrand}, {Applications and trends in wireless acoustic sensor networks: A
  signal processing perspective}, in: Proc. 18th IEEE SCVT, 2011, pp. 1--6.
\newblock \href {https://doi.org/10.1109/SCVT.2011.6101302}
  {\path{doi:10.1109/SCVT.2011.6101302}}.

\bibitem{zhang2018microphone}
J.~Zhang, S.~P. Chepuri, R.~C. Hendriks, R.~Heusdens, {Microphone Subset
  Selection for MVDR Beamformer Based Noise Reduction}, IEEE/ACM Trans. Audio,
  Speech, Lang. Process. 26~(3) (2018) 550--563.
\newblock \href {https://doi.org/10.1109/TASLP.2017.2786544}
  {\path{doi:10.1109/TASLP.2017.2786544}}.

\bibitem{mehanna2013joint}
O.~Mehanna, N.~D. Sidiropoulos, G.~B. Giannakis, {Joint Multicast Beamforming
  and Antenna Selection}, IEEE Trans. Signal Process. 61~(10) (2013)
  2660--2674.
\newblock \href {https://doi.org/10.1109/tsp.2013.2252167}
  {\path{doi:10.1109/tsp.2013.2252167}}.

\bibitem{hamza2019hybrid}
S.~A. Hamza, M.~G. Amin, {Hybrid Sparse Array Beamforming Design for General
  Rank Signal Models}, IEEE Trans. Signal Process. 67~(24) (2019) 6215--6226.
\newblock \href {https://doi.org/10.1109/TSP.2019.2952052}
  {\path{doi:10.1109/TSP.2019.2952052}}.

\bibitem{geirnaert2020fast}
S.~{Geirnaert}, T.~{Francart}, A.~{Bertrand}, {Fast EEG-based decoding of the
  directional focus of auditory attention using common spatial patterns}, IEEE
  Trans. Biomed. Eng. 68~(5) (2021) 1557--1568.
\newblock \href {https://doi.org/10.1109/TBME.2020.3033446}
  {\path{doi:10.1109/TBME.2020.3033446}}.

\bibitem{blankertz2007optimizing}
B.~Blankertz, R.~Tomioka, S.~Lemm, M.~Kawanabe, K.-R. Muller, {Optimizing
  spatial filters for robust EEG single-trial analysis}, IEEE Signal Process.
  Mag. 25~(1) (2007) 41--56.
\newblock \href {https://doi.org/10.1109/MSP.2008.4408441}
  {\path{doi:10.1109/MSP.2008.4408441}}.

\bibitem{dash1997feature}
M.~Dash, H.~Liu, {Feature Selection for Classification}, Intell. Data Anal.
  1~(1) (1997) 131--156.
\newblock \href {https://doi.org/10.1016/S1088-467X(97)00008-5}
  {\path{doi:10.1016/S1088-467X(97)00008-5}}.

\bibitem{joshi2009sensor}
S.~Joshi, S.~Boyd, {Sensor selection via convex optimization}, IEEE Trans.
  Signal Process. 57~(2) (2009) 451--462.
\newblock \href {https://doi.org/10.1109/TSP.2008.2007095}
  {\path{doi:10.1109/TSP.2008.2007095}}.

\bibitem{vanveen1988beamforming}
B.~Van~Veen, K.~Buckley, Beamforming: a versatile approach to spatial
  filtering, IEEE ASSP Mag. 5~(2) (1988) 4--24.
\newblock \href {https://doi.org/10.1109/53.665} {\path{doi:10.1109/53.665}}.

\bibitem{yan2006trace}
S.~Yan, X.~Tang, Trace quotient problems revisited, in: A.~Leonardis,
  H.~Bischof, A.~Pinz (Eds.), Computer Vision -- ECCV 2006, Springer Berlin
  Heidelberg, Berlin, Heidelberg, 2006, pp. 232--244.
\newblock \href {https://doi.org/10.1007/11744047_18}
  {\path{doi:10.1007/11744047_18}}.

\bibitem{wang2014reconfigurable}
X.~Wang, E.~Aboutanios, M.~Trinkle, M.~G. Amin, {Reconfigurable Adaptive Array
  Beamforming by Antenna Selection}, IEEE Trans. Signal Process. 62~(9) (2014)
  2385--2396.
\newblock \href {https://doi.org/10.1109/TSP.2014.2312332}
  {\path{doi:10.1109/TSP.2014.2312332}}.

\bibitem{wang2015adaptive}
X.~Wang, E.~Aboutanios, M.~G. Amin, {Adaptive Array Thinning for Enhanced DOA
  Estimation}, IEEE Signal Process. Lett. 22~(7) (2015) 799--803.
\newblock \href {https://doi.org/10.1109/LSP.2014.2370632}
  {\path{doi:10.1109/LSP.2014.2370632}}.

\bibitem{luo2010semidefinite}
Z.-Q. Luo, W.-K. Ma, A.~M.-C. So, Y.~Ye, S.~Zhang, {Semidefinite Relaxation of
  Quadratic Optimization Problems}, IEEE Signal Process. Mag. 27~(3) (2010)
  20--34.
\newblock \href {https://doi.org/10.1109/MSP.2010.936019}
  {\path{doi:10.1109/MSP.2010.936019}}.

\bibitem{candes2008enhancing}
E.~J. Cand{\`{e}}s, M.~B. Wakin, S.~P. Boyd, {Enhancing Sparsity by Reweighted
  $\ell_1$ Minimization}, J. Fourier Anal. Appl. 14~(5-6) (2008) 877--905.
\newblock \href {https://doi.org/10.1007/s00041-008-9045-x}
  {\path{doi:10.1007/s00041-008-9045-x}}.

\bibitem{grant2014cvx}
M.~Grant, S.~Boyd, {CVX: Matlab Software for Disciplined Convex Programming,
  version 2.2}, \url{http://cvxr.com/cvx} (2020).

\bibitem{grant2008graph}
M.~Grant, S.~Boyd, Graph implementations for nonsmooth convex programs, in:
  V.~Blondel, S.~Boyd, H.~Kimura (Eds.), Recent Advances in Learning and
  Control, Lecture Notes in Control and Information Sciences, Springer-Verlag
  Limited, 2008, pp. 95--110, \url{http://stanford.edu/\~boyd/graph\_dcp.html}.

\bibitem{mosek2019}
{MOSEK ApS}, \href{http://docs.mosek.com/9.1/toolbox/index.html}{The MOSEK
  optimization toolbox for MATLAB manual. Version 9.1.9} (2019).
\newline\urlprefix\url{http://docs.mosek.com/9.1/toolbox/index.html}

\bibitem{golub2000eigenvalue}
G.~H. Golub, H.~A. {Van Der Vorst}, {Eigenvalue computation in the 20th
  century}, J. Comput. Appl. Math. 123~(1-2) (2000) 35--65.
\newblock \href {https://doi.org/10.1016/S0377-0427(00)00413-1}
  {\path{doi:10.1016/S0377-0427(00)00413-1}}.

\bibitem{qi2021spatiotemporal}
F.~{Qi}, W.~{Wu}, Z.~L. {Yu}, Z.~{Gu}, Z.~{Wen}, T.~{Yu}, Y.~{Li},
  {Spatiotemporal-Filtering-Based Channel Selection for Single-Trial EEG
  Classification}, IEEE Trans. Cybern. 51~(2) (2021) 558--567.
\newblock \href {https://doi.org/10.1109/TCYB.2019.2963709}
  {\path{doi:10.1109/TCYB.2019.2963709}}.

\bibitem{meng2009automated}
J.~{Meng}, G.~{Liu}, G.~{Huang}, X.~{Zhu}, {Automated selecting subset of
  channels based on CSP in motor imagery brain-computer interface system}, in:
  Proc. of IEEE Int. Conf. ROBIO, 2009, pp. 2290--2294.
\newblock \href {https://doi.org/10.1109/ROBIO.2009.5420462}
  {\path{doi:10.1109/ROBIO.2009.5420462}}.

\bibitem{arvaneh2011optimizing}
M.~{Arvaneh}, C.~{Guan}, K.~K. {Ang}, C.~{Quek}, {Optimizing the Channel
  Selection and Classification Accuracy in EEG-Based BCI}, IEEE Trans. Biomed.
  Eng. 58~(6) (2011) 1865--1873.
\newblock \href {https://doi.org/10.1109/TBME.2011.2131142}
  {\path{doi:10.1109/TBME.2011.2131142}}.

\bibitem{onaran2013sparse}
I.~Onaran, N.~F. Ince, A.~E. Cetin, {Sparse spatial filter via a novel
  objective function minimization with smooth $\ell_1$ regularization}, Biomed.
  Signal Process. Control 8~(3) (2013) 282--288.
\newblock \href {https://doi.org/https://doi.org/10.1016/j.bspc.2012.10.003}
  {\path{doi:https://doi.org/10.1016/j.bspc.2012.10.003}}.

\bibitem{galecki2013linear}
A.~{Gałecki}, T.~{Burzykowski}, {Linear Mixed-Effects Models Using R: A
  Step-by-Step Approach}, Springer Texts in Statistics, Springer-Verlag New
  York, 2013.
\newblock \href {https://doi.org/10.1007/978-1-4614-3900-4}
  {\path{doi:10.1007/978-1-4614-3900-4}}.

\end{thebibliography}
\end{document}